\documentclass[twoside]{IEEEtran}

\newif\ifhyperref\hyperreftrue

\usepackage{mathrsfs}
\usepackage{amsmath}
\usepackage{cases}
\usepackage{amssymb}
\usepackage{graphicx}
\usepackage{url}

\usepackage{enumerate}
\makeatletter
\def\@@enum@[#1]{
  \@enLab{}\let\@enThe\@enQmark
  \@enloop#1\@enum@
  \ifx\@enThe\@enQmark\@warning{The counter will not be printed.%
   ^^J\space\@spaces\@spaces\@spaces The label is: \the\@enLab}\fi
  \expandafter\edef\csname label\@enumctr\endcsname{\the\@enLab}%
  \expandafter\let\csname the\@enumctr\endcsname\@enThe
  \csname c@\@enumctr\endcsname7
  \@enum@}
\makeatother

\usepackage{calc}

\newlength{\myleftmargini}

\newsavebox{\stepspaceonedot}\savebox{\stepspaceonedot}
{\hspace*{0em}Step 1.{}\hskip\labelsep}
\newlength{\myitemindentstep} 
\newsavebox{\casespaceonecolon}\savebox{\casespaceonecolon}
{\hspace*{0em}Case 1:{}\hskip\labelsep}
\newlength{\myitemindentcase} 
\newsavebox{\oneacolon}\savebox{\oneacolon}
{\hspace*{0em}1(a):{}\hskip\labelsep}
\newlength{\myitemindentonea} 
\newsavebox{\oneaiiicolon}\savebox{\oneaiiicolon}
{\hspace*{0em}1(a)iii:{}\hskip\labelsep}
\newlength{\myitemindentoneaiii} 
\newsavebox{\conditionone}\savebox{\conditionone}
{\hspace*{0em}(1)\hskip\labelsep}

\usepackage{xspace}

\ifhyperref%
\usepackage{color,svgcolor}
\usepackage[%
breaklinks%
,pdfa
,colorlinks%
,linkcolor=red
,anchorcolor=mediumblue%
,citecolor=mediumblue%
,bookmarks=false%
]{hyperref}
\usepackage{breakurl}
\fi

\usepackage{theorem}

\usepackage{longtable}
\usepackage{supertabular}

\def\citeP{\cite}

\newcommand{\EKhref}[2]{URL: \url{#1}}

\newtheorem{theorem}{Theorem}
\newtheorem{prop}[theorem]{Proposition}

\newtheorem{lemma}[theorem]{Lemma}

{\theorembodyfont{\upshape}
\newtheorem{algorithm}
{Algorithm}
}

{\theorembodyfont{\upshape}
\newtheorem{remark}{Remark}
}

{\theorembodyfont{\upshape}
\newtheorem{example}
{Example}
}

{Problem}

{\theoremstyle{plain}

}

\makeatletter
\def\@yproof[#1]{\@proof{ #1}}
\def\@proof#1{\begin{trivlist}\item[]{\em Proof#1.}}

\makeatother

\newcommand{\chebT}{Cheby\-\hbox{shev-1}\xspace}

\newcommand{\CC}{\mathbb{C}}
\newcommand{\RR}{\mathbb{R}}
\newcommand{\ZZ}{\mathbb{Z}}

\newcommand{\KK}{\mathsf{K}} 

\newcommand{\defequal}{%
\begin{tabular}[b]{@{}c@{}}
\text{\hbox to 0pt{\hss\footnotesize\upshape def\hss}}
\\[-1.35ex]
${}={}$
\end{tabular}}

\newcommand{\ruleaboveheader}[1]
{\parbox[t][0cm][t]{0cm}{%
\hspace*{0cm}
\parbox[b][0cm][t]{0cm}{%
\vspace*{#1}
\rule{\columnwidth}{0.5pt}
\par}\par}\par\nobreak\vspace{-2ex}\noindent}

\newcommand{\ruleaboveonelineheader}{\ruleaboveheader{-5ex}}
\newcommand{\ruleabovetwolineheader}{\ruleaboveheader{-10.4ex}}

\newcommand{\rulebelowtext}[1]
{\hfill
\parbox[t][0cm][t]{0cm}{%
\hspace*{-\columnwidth}
\parbox[b][0cm][t]{0cm}{%
\vspace*{#1}
\rule{\columnwidth}{0.5pt}
\par}\par}}

\newcommand{\ourtriang}{\raise2.5pt\hbox{\tiny$\blacktriangleright$\ }}

\newsavebox{\onedotspace}\savebox{\onedotspace}{\hspace*{0em}1.{}\hskip\labelsep}%

\newcommand{\fsupbr}[1]{f^{\,[ #1 ]}} 
\newcommand{\eval}{\hat{a}}
\newcommand{\sHankel}{H}
\newcommand{\sfoldmatrix}{G}
\newcommand{\numofblocks}{\theta}
\newcommand{\numofevals}{N}

\renewcommand{\IEEEQED}{\IEEEQEDopen}
\renewcommand{\IEEEiedlistdecl}{\settowidth{\labelwidth}{}}
\renewcommand{\IEEEiedlistdecl}{\settowidth{\labelsep}{}}

\begin{document}

\title{Sparse Interpolation With Errors in {Chebyshev} Basis
Beyond Redundant-Block Decoding
}

\author{
Erich~L.~Kaltofen and Zhi-Hong~Yang
\thanks{Erich~L.~Kaltofen and Zhi-Hong Yang are 
with the
Department of Mathematics, North Carolina State University,
Raleigh, North Carolina 27695-8205, USA, and also with the
Department of Computer Science, Duke University,
Durham, North Carolina 27708-0129, USA
(email: kaltofen@ncsu.edu,\allowbreak kaltofen@cs.duke.edu,
zhihongyang2020@outlook.com). 
}
\thanks{
This research was supported by
the National Science Foundation
under Grant CCF-1717100 (Kaltofen and Yang).
}
\thanks{ 
Copyright (c) 2020 
IEEE.
Personal use of this material is permitted.  Permission from IEEE must
be obtained for all other uses, in any current or future media, including
reprinting/republishing this material for advertising or promotional purposes,
creating new collective works, for resale or redistribution to servers or lists, or
reuse of any copyrighted component of this work in other works.
}
}

\maketitle

\begin{abstract}
\boldmath 
\noindent
We present sparse interpolation algorithms for recovering 
a polynomial with $\le B$ terms from $N$ evaluations at distinct
values for the variable when $\le E$ of the evaluations
can be erroneous.  Our algorithms perform exact arithmetic 
in the field of scalars~$\KK$ 
and the terms can be standard powers of the
variable or Chebyshev polynomials, in which case the characteristic
of $\KK$ is $\ne 2$.  Our algorithms return a list of valid
sparse interpolants for the $N$ support points and run in
polynomial-time.  
For standard power basis our algorithms 
sample at $N = \lfloor \frac{4}{3} E + 2 \rfloor B$ points, 
which are fewer points than $N = 2(E+1)B - 1$ 
given by  Kaltofen and Pernet in 2014. 
For Chebyshev basis our algorithms sample 
at $N = \lfloor \frac{3}{2} E + 2 \rfloor B$ points, 
which are also fewer than the
number of 
points required by the algorithm 
given by Arnold and Kaltofen in 2015,
which has $N = 74 \lfloor \frac{E}{13} + 1 \rfloor$ 
for $B = 3$ and $E \ge 222$. 
Our method shows how to correct $2$ errors in a block of $4B$ points
for standard basis and how to correct $1$ error in a block
of $3B$ points for Chebyshev Basis.
\end{abstract}


\section{Introduction}\label{sec:intro}

\IEEEPARstart{L}{et} 
$f(x)$ be a polynomial with coefficients from a field $\KK$
(of characteristic $\ne 2$),
\begin{multline}\label{eq:fx}
f(x) = \sum_{j=1}^t c_j T_{\delta_j}(x)\in\KK[x],\;
\\0 \le \delta_1 < \delta_2 < \cdots < \delta_t = \deg(f),
\forall j, 1\le j\le t\colon c_j \ne 0,
\end{multline}
where $T_d(x)$ is the Chebyshev Polynomial of the First Kind
(of degree $d$ for $d\ge 0$), defined by the recurrence
\begin{equation}\label{def:matpow}
\begin{bmatrix} T_d(x) \\ T_{d+1}(x) \end{bmatrix} =
\begin{bmatrix} 0 & 1\\ -1 & 2x \end{bmatrix}^d
\begin{bmatrix} 1 \\ x \end{bmatrix}
\quad\text{ for $d\in\ZZ$.}
\end{equation}
We say that $f(x)$ is \chebT $t$-sparse.  We wish to compute
the term degrees $\delta_j$ and the coefficients $c_j$ from
values of $a_i = f(\zeta_i)$ for $i=1,2,\ldots$, where 
the distinct arguments $\zeta_i\in\KK$ can
be chosen by the algorithms; the latter is the setting
of Prony-like sparse interpolation methods.
Our objective is to interpolate with a number of points
that is proportional to the sparsity~$t$ of~$f$. 
The algorithms have as input
an upper bound $B \ge t$ for the sparsity, for otherwise
the zero polynomial (of sparsity~$0$) is indistinguishable from
$f(x) = \prod_i(x-\zeta_i)$ at $\le\deg(f)$ evaluation points $a_i = 0$.
The algorithms by \citeP{LaSa95,ArKa15,IKY18}, based on
Prony-like interpolation \citeP{deProny,B-OTi88,KL03}, can
interpolate $f(x)$ (see \eqref{eq:fx}) 
from $2B$ values at points
$\zeta_i = T_i(\beta)=(\omega^i+1/\omega^i)/2$ 
for $i=0,1,\ldots,2B-1$ where
$\beta=(\omega+1/\omega)/2$ with $\omega\in\KK$ such that
$\omega^{\delta_j} \ne \omega^{\delta_k}$ for all $1 \le j < k \le t$.
Like Prony's original algorithm, 
our algorithms utilize an algorithm
for computing roots in $\KK$ of polynomials with coefficients in $\KK$ 
and logarithms to base~$\omega${}. More precisely, one utilizes
an algorithm that on input $\omega$ and $\omega^d$ for
an integer $d\in\ZZ$ computes $d$, possibly modulo the finite
multiplicative order $\eta$ of $\omega$ ($\omega^\eta = 1$ minimally) \citeP{ImKa18}.
We note that in \citeP{ArKa15} we show that one may instead use
the odd-indexed argument $T_{2i+1}(\beta)$ for $i=0,1,...,2B-1$,
provided $\omega^{2\delta_j+1} \ne \omega^{2\delta_k+1}$
for all $1 \le j < k \le t$.

\begin{figure}[htbp]
\centering
\includegraphics[width=0.5\textwidth]{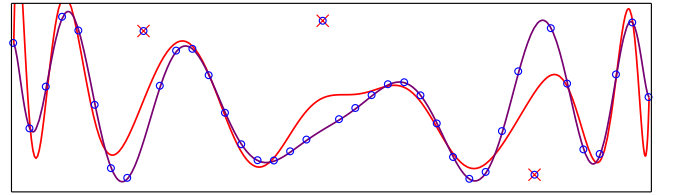}
\caption{Sparse \chebT polynomial fit after removing $3$ errors vs.  
polynomial least squares fit}
\label{fig:overfit_cheb_lsq}
\end{figure}

Here we consider the case when the evaluations $a_i$, which we think
of being computed by probing a black box that evaluates $f$,
can have sporadic errors.  We write $\hat a_i$ for the black box values, which
at some unknown indices $\ell$ can have $\hat a_\ell \ne a_\ell$.
In the plot in Fig.~\ref{fig:overfit_cheb_lsq}, 
which is for the
range $-1\leq x\leq 1$, the purple function is $T_{15}(x) - 2T_{11}(x)
+ T_2(x)$ that fits $37$ of the $40$ values, while the red model is a
polynomial least squares fit of degree $\leq 19$. 
The red function captures 3 possible outliers,
resulting in a model which has a lower accuracy on the remaining $37$ data points.

We shall assume that we have an upper bound $E$ for the number of
errors on a batch of $N$ evaluations.
Therefore our sequence of black box calls has a non-stochastic
error rate $\le E/N$.
We shall also assume that
the black box for $f$ does not return stochastic errors, meaning that
if $\hat a \ne f(\zeta)$ then a second evaluation of the black box
at $\zeta$ produces the same erroneous $\hat a$.
Furthermore, we perform list-interpolation which produces a
valid list of sparse interpolants for the black box values with errors,
analogously to list-decoding error correcting codes.
We restrict to algorithms that run in polynomial time in $B$ and $E$
($N$ is computed by the algorithms),
which limits the list length 
to be 
polynomial in $B$ and $E$.

A simple sparse list-interpolation algorithm with errors evaluates
$E+1$ blocks of $2B$ arguments, which produce $N = (E+1)2B$ black box values
$\hat a_{i,\sigma}$ at the arguments
\begin{equation}\label{eq:block_args}
\left.
\begin{array}{llll}
T_1(\beta_1),&T_3(\beta_1),&\ldots,&T_{4B-1}(\beta_1), 
\\
T_1(\beta_2),&T_3(\beta_2),&\ldots,&T_{4B-1}(\beta_2), 
\\
\vdots&\vdots & & \vdots
\\
T_1(\beta_{E+1}),&T_3(\beta_{E+1}),&\ldots,&T_{4B-1}(\beta_{E+1}), 
\end{array}
\right\} E+1
\end{equation}
where $\beta_\sigma = (\omega_\sigma + 1/\omega_\sigma)/2$
and where the arguments in (\ref{eq:block_args}) are selected distinct:
$T_{2i+1}(\beta_\sigma) \neq T_{2m+1}(\beta_\tau)$
for $i \ne m$ and $\sigma\ne\tau$
($\Longleftrightarrow \omega_\sigma^{2i+1} \ne \omega_\tau^{2m+1}$).
If we have for all $\omega_\sigma$ distinct term values
$\omega_\sigma^{\delta_j} \ne \omega_\sigma^{\delta_k}$ ($j\ne k$)
then the algorithm in \citeP{ArKa15} can recover $f$ from
those lines 
in (\ref{eq:block_args}) at which the black box does not evaluate
to an error, because we assume $\le E$ errors there is such a block
of good arguments/values.  Other blocks with errors may lead to a different
$t$-sparse \chebT interpolant with $t\le B$. 
The goal is to recover $f$ (and possible
other sparse interpolants with $\le E$ errors) from $N < (E+1)2B$ evaluations.

In \citeP{ArKa15} we give algorithms for the following bounds $B,E\colon$
\begin{equation}\label{eq:ArKa15}
\begin{tabular}{%
@{}l@{}%
}
$B=1\colon\forall E \ge 57\colon$
\\ $\quad$ $\displaystyle N = 23 \lfloor \frac{E}{14}+1 \rfloor < 2 (E+1) = 2B(E+1);$
 $\displaystyle\frac{23}{14} \le 1.65,$
\\[2ex]
$B=2\colon\forall E \ge 86\colon$
\\
$\quad$ $\displaystyle N = 43 \lfloor \frac{E}{12}+1 \rfloor < 4 (E+1) = 2B(E+1);$
 $\displaystyle\frac{43}{12} \le 3.59,$
\\[2ex]
$B=3\colon\forall E \ge 222\colon$
\\
$\quad$ $\displaystyle N = 74 \lfloor\frac{E}{13} + 1 \rfloor < 6
(E+1) = 2B(E+1);$ 
 $\displaystyle\frac{74}{13} \le 5.70.$
\end{tabular}
\end{equation}
The evaluation counts (\ref{eq:ArKa15}) are derived by using
the method of \citeP{KaPe14}: subsampling at all subsequences
$x \gets T_{r+is}(\beta)$ of arguments
whose indices are arithmetic progressions to locate a subsequence
without an error.  The counts (\ref{eq:ArKa15}) are established by
explicitly computed lengths for the Erd\H{o}s-Tur\'{a}n Problem for
arithmetic progressions of length 
$3B$ when $B=1,\;2,\;3$. 
For an arbitrary positive integer $B$, 
Gowers's 2001 effective estimates 
\citeP[Theorem~1.3]{Gow01} 
for Szemeredi's proof 
of the Erd\H{o}s-Tur\'{a}n Conjecture allow
us to compute a lower bound 
for $E$ when subsampling requires fewer than $2B(E+1)$ values, 
but the lower bound is quintuply exponential in $B${}. 
Here we give an algorithm that recovers $f$ (and possible
other sparse interpolants) {\itshape for all} $B\ge 1,E\ge 1$ bounds from
\begin{equation}\label{eq:cheb_res}
\numofevals = \left\lfloor \frac{3}{2}E + 2\right\rfloor B 
\end{equation}
evaluations with $\le E$ errors.
Our new algorithm uses
fewer evaluations than (\ref{eq:ArKa15}). 
We show that
one can list-interpolate from $3B$ points correcting a single error,
which with blocking yields (\ref{eq:cheb_res}).
We correct one error from $3B$ points by deriving a non-trivial
univariate polynomial for the value as a variable in each
possible position.

Our technique applies to Prony's original problem of
interpolating a $t$-sparse polynomial with $t \le B$ 
in power basis $1,x,x^2,\ldots$ in the presence
of erroneous points.  In \citeP[Lemma~2]{KaPe14} it was shown that
from $(E+1)2B-1$ points one can correct $\le E$ errors.
Here we show that
\begin{equation}\label{eq:prony_res}
\numofevals = \left\lfloor\frac{4}{3}E + 2\right\rfloor B
\end{equation}
points suffice to correct $\le E$ errors.
The count (\ref{eq:prony_res}) is 
achieved by correcting
$\le 2$ errors from $4B$ points and blocking.
We correct $2$ errors at $4B$ points by deriving
a bivariate Pham system for variables in place of
the values in all possible error locations, which yields a
bounded number of possible value pairs among which are the
actual values.  We note
that for $E = 2$ the count $4B$ is smaller than the values $n_{2B,2}$
in \citeP[Table~1]{KaPe14}, which are the counts for having
a clean arithmetic progression of length $2B$ in the presence of $2$ errors.

Our algorithms for interpolating sparse polynomials in 
power basis (or Laurent polynomials) with
errors, can tolerate a higher error rate $E/N$ 
than the existing algorithms
in \citeP{CKP12} and \citeP{KaPe14}. 
For correcting $E$ errors, the algorithm in \citeP{CKP12} 
uses redundant-block decoding which requires 
$N=2B(E+1)$ points, and the algorithm in \citeP{KaPe14} uses 
subsampling which
is shown by an explicit analysis of the arising Erd\H{o}s-Tur\'{a}n Problem to require no more than 
$N=2B(E+1)-1$ points. 
That is the best we have been able to do {\itshape for all} $B$ and $E$ using subsampling. 
In this paper, 
we use a different technique.
We correct one error in a block of $3B$ points, or correct
$2$ errors in a block of $4B$ points, by replacing possible 
errors with symbols, and then solve for the symbols to
obtain the actual values; next with redundant-block decoding,
we can correct $E$ errors
from $N=\left\lfloor\frac{4}{3}E + 2\right\rfloor B$ points,
for all $B$ and $E$.  
Since Chebyshev polynomials can be transformed
into Laurent polynomials \eqref{eqn:sparse_Laurent},
we first discuss our new 
algorithms for Laurent polynomials 
in Section~\ref{sec:power_basis},
and then apply 
the same technique for Chebyshev bases. 

Finally we note that our sparse list-interpolation algorithms are
interpolation algorithms over the reals $\KK = \RR$
if $\omega_\sigma > 1$ (or $\omega_\sigma > 0$ when $f$ is in power
basis) and $\numofevals \geq 2B+2E$,
that is, there will only
be a single sparse interpolant computed by our algorithms.
Uniqueness is a consequence of Descartes's Rule of Signs and its
generalization to polynomials in orthogonal bases by Obrechkoff's
Theorem of 1918 \citeP{DiRa2009} 
(see also Corollary~2 in \citeP{KaPe14} and
Corollary~2.4 in \citeP{ArKa15}).
Over fields with roots of unity, the sparse list-interpolation problem
for the  power bases 
with $< (2E+1)2B$ points can have more than a single
$B$-sparse solution \citeP[Theorem~3]{KaPe14}, 
which is also true for the 
\chebT basis 
as shown by 
Example~\ref{example:cheb_unique_decoding}. 

\section{Sparse Interpolation in Standard Power Basis with 
Error Correction}
\label{sec:power_basis}
\subsection{Correcting One Error}
\label{subsec:one_error_power} 

Let $ \KK $ be a
field of scalars. 
Let $ f(x)\in \KK[x,x^{-1}] $ be 
a sparse univariate Laurent polynomial 
represented by a black box
and it is equal to:
\begin{multline}\label{eqn:f_powerbasis} 
	f(x)=\sum_{j=1}^{t} c_j x^{\delta_j},\;
	\delta_1 < \delta_2 < \cdots < \delta_t = \deg(f),\\
	\forall  j, 1\le j\le t\colon c_j \neq 0.
\end{multline} 
We assume that the black box for $f$ returns the same value
when probed multiple times at the same input.
Let $B$ be an upper bound on the sparsity of $f(x)$
and $D\geq |\delta_j|$ for all $1\leq j\leq t$.
Choose a point $\omega\in \KK\setminus\{0\} $ such that: 
{\setlength{\leftmargini}{\wd\conditionone} 
\begin{enumerate}[(1)] 
\item $\omega$ has order $\geq 2D+1$, meaning that 
$\forall \eta, 1 \le \eta \le 2D\colon  \omega^{\eta}\ne 1 $.
\item  $ \omega^{i_1}\neq \omega^{i_2} $
for all $ 1\leq i_1 < i_2\leq 3B $. 
\end{enumerate}
}
The first condition is
an input specification of the Integer Logarithm Algorithm (see 
Algorithm~\ref{alg:intlog})
that computes $\delta_j$ from
$\omega^{\delta_j}$. 
The second condition guarantees that the inputs 
probed at the black box are distinct so that 
we don't get the same error at different
locations.

For $ i=1,2,\ldots, 3B
$, let $ \eval_i $ be the output of the black box for $ f$ probed at
input $ \omega^i $. Assume there is at most one error 
in the evaluations, that is, there
exists $\leq 1$ index $ i\in \{1,2, \ldots, 3B \} $ such 
that $ \eval_i \neq f(\omega^i) $. We present an algorithm 
to compute a list of sparse polynomials which contains $ f $. 


For $r=1,\ldots,B$, 
let $\sHankel_r \in \KK^{(B+1)\times(B+1)}$ be
the following Hankel matrix: 
\begin{equation}\label{eqn:Hankel} 
	\sHankel_r = 
	\begin{bmatrix}
	\eval_r        & \eval_{r+1}   & \cdots &
	\eval_{r+B-1}  & \eval_{r+B}   \\ 
	\eval_{r+1}    & \eval_{r+2}   & \cdots &  
	\eval_{r+B}    & \eval_{r+B+1} \\ 
	\vdots         & \vdots        & \vdots &  
	\vdots         & \vdots        \\ 
	\eval_{r+B-1}  & \eval_{r+B}   & \cdots & 
	\eval_{r+2B-2} & \eval_{r+2B-1}\\
	\eval_{r+B}    & \eval_{r+B+1} & \cdots & 
	\eval_{r+2B-1} & \eval_{r+2B} 
	\end{bmatrix}
\end{equation}

Let $\ell$ be the error location, i.e., $\eval_{\ell}\neq
f(\omega^{\ell})$. There are three cases to be considered:

{\setlength{\leftmargini}{\myleftmargini}
\begin{enumerate}[{Case} 1:]
\setlength{\itemindent}{\myitemindentcase}

\item\label{case:1error_at_head_power}
$1\leq \ell\leq B$;

\item\label{case:1error_at_middle_power}
$B+1\leq \ell \leq 2B$;
 
\item\label{case:1error_at_tail_power} 
$2B+1\leq \ell\leq 3B$.
\end{enumerate}
}

For Case~\ref{case:1error_at_head_power} and 
Case~\ref{case:1error_at_tail_power}, we can use Prony's algorithm
(see Algorithm~\ref{alg:prony}) to
recover $f(x)$ from a consecutive sequence of length $2B$: either
$(\eval_1,\eval_2, \ldots, \eval_{2B}  )$ or 
$(\eval_{B+1},\eval_{B+2}, \ldots, \eval_{3B} )$.
To deal with Case \ref{case:1error_at_middle_power}, we replace the
erroneous value $\eval_{\ell}$ by a symbol $\alpha$. Then
the determinant the Hankel matrix $\sHankel_{\ell-B} $
(see \eqref{eqn:Hankel}) is
univariate polynomial of degree $B+1$ in $\alpha$.
By Prony/Blahut/Ben-Or/Tiwari Theorem \citeP{deProny,Blahut1983, B-OTi88},
$ (f(\omega^{i}))_{i\geq 0} $ 
is a linearly generated sequence
and its minimal generator has degree $\leq B$. Therefore
$f(\omega^{\ell})$ is a solution of the equation:
\begin{equation}\label{eqn:det_1error_power}
	\det(\sHankel_{\ell-B}) = 0.
\end{equation} 
By solving the equation \eqref{eqn:det_1error_power}, we
obtain a list of candidates 
$\{\xi_1,\ldots,\xi_b  \}$ 
for the correct value
$f(\omega^{\ell})$. For each candidate $\xi_k (1\leq k\leq
b)$, we substitute $\eval_{\ell}$ by $\xi_k$ in the sequence
$( \eval_{B+1},\eval_{B+2}, \ldots,\eval_{2B} )$ and try Prony's
algorithm on the updated sequence 
$(\eval_1, \eval_2, \ldots, \eval_{2B} )$,
which gives us a list of sparse polynomials 
containing $f(x)$. 
The process of correcting one error from $3B$ evaluations 
is illustrated by the following example.

\begin{example}
Assume that we are given $B=3$. With $3B=9$ evaluations
$\eval_1,\eval_2,\ldots,\eval_9$ obtained from the black box
for $f$ at inputs $\omega, \omega^2,\ldots, \omega^9$, 
we have the following $6\times 4$ matrix:
\begin{equation*}
\sHankel =
\begin{bmatrix}
\eval_1 & \eval_2 & \eval_3 & \eval_4 \\
\eval_2 & \eval_3 & \eval_4 & \eval_5 \\
\eval_3 & \eval_4 & \eval_5 & \eval_6 \\
\eval_4 & \eval_5 & \eval_6 & \eval_7 \\
\eval_5 & \eval_6 & \eval_7 & \eval_8 \\
\eval_6 & \eval_7 & \eval_8 & \eval_9 \\
\end{bmatrix}\in \KK^{6\times 4}
\end{equation*}
For $r=1,2,3$, the matrices $\sHankel_r$ (see \eqref{eqn:Hankel}) 
are $4\times 4$
submatrices of $\sHankel$:
\begin{gather*}
\sHankel_1 =
\begin{bmatrix}
\eval_1 & \eval_2 & \eval_3 & \eval_4 \\
\eval_2 & \eval_3 & \eval_4 & \eval_5 \\
\eval_3 & \eval_4 & \eval_5 & \eval_6 \\
\eval_4 & \eval_5 & \eval_6 & \eval_7 \\
\end{bmatrix},\
\sHankel_2 =
\begin{bmatrix}
\eval_2 & \eval_3 & \eval_4 & \eval_5 \\
\eval_3 & \eval_4 & \eval_5 & \eval_6 \\
\eval_4 & \eval_5 & \eval_6 & \eval_7 \\
\eval_5 & \eval_6 & \eval_7 & \eval_8 \\
\end{bmatrix},\\
\sHankel_3 =
\begin{bmatrix}
\eval_3 & \eval_4 & \eval_5 & \eval_6 \\
\eval_4 & \eval_5 & \eval_6 & \eval_7 \\
\eval_5 & \eval_6 & \eval_7 & \eval_8 \\
\eval_6 & \eval_7 & \eval_8 & \eval_9 \\
\end{bmatrix}.
\end{gather*}
Suppose there is one error $\eval_{\ell}\neq f(\omega^{\ell})$ 
in these $3B$ evaluations. We recover $f(x)$ by the
following steps.

{\setlength{\leftmargini}{\myleftmargini}

\begin{enumerate}[Step 1:]
\setlength{\itemindent}{\myitemindentstep}

\item Try to recover $f(x)$ from $(\eval_1,\eval_2,\ldots,\eval_6)$
and $(\eval_4,\eval_5,\ldots,\eval_9)$ by Prony's algorithm;
$f(x)$ will be returned if $\ell \in \{7,8,9\}$ or $ \ell \in \{ 1,2,3\}$.
\item For $\ell\in \{4,5,6\}$, substitute $\eval_{\ell}$ by
$\alpha$, then $\det(\sHankel_{\ell-3})$ is a univariate
polynomial of degree $4$ in $\alpha$ and $f(\omega^{\ell})$
is a root of $\det(\sHankel_{\ell-3})$. Compute the roots
$\{\xi_k\}_{k\geq 1}$ of $\det(\sHankel_{\ell-3})$.
For each root $\xi_k$, replace $\eval_{\ell}$ by $\xi_k$
and check if the matrix $\sHankel$ has rank $\leq 3$.
If yes, then use Prony's algorithm (see Algorithm~\ref{alg:prony}) 
on the updated
sequence $(\eval_1,\eval_2,\ldots,\eval_6)$. As $f(\omega^{\ell})$ is equal to some $\xi_k$, this step will recover $f(x)$ in
case that $\ell\in \{4,5,6\}$.
\end{enumerate}
}
\end{example}



For computing the term degrees $\delta_j$ of $f$, we need 
an integer logarithm algorithm, 
please see Algorithm~\ref{alg:intlog}  
{\itshape Integer Logarithm Algorithm} for the input and output specifications.  

\vspace{-2ex}
\begin{algorithm}
\vspace*{1ex}
{\itshape Integer Logarithm Algorithm}
\newline
\ruleaboveonelineheader
\label{alg:intlog}
{\itshape Input:\/}
\begin{itemize}
\item[\ourtriang]
An upper bound $D\in \ZZ_{>0}$.
\item[\ourtriang]
$\omega\in \KK \setminus \{0\}$ and has order $\geq 2D+1$,
meaning that $\forall \eta\geq 1, 
\omega^{\eta}=1 $ $\Rightarrow$  $\eta\geq 2D+1$.
\item[\ourtriang]
$\rho\in \KK \setminus \{0\}$.
\end{itemize}
{\itshape Output:\/}
\begin{itemize}
\item[\ourtriang]
Either $\delta \in \ZZ$ with $|\delta|\leq D$ and $\omega^{\delta} = \rho$,
\item[\ourtriang]
or FAIL.
\end{itemize}
\rulebelowtext{-2ex}
\end{algorithm}

\vspace{-3ex}

We describe the subroutine which we call Try Prony's
algorithm. This subroutine will be frequently used in our
main algorithms,
please see Algorithm~\ref{alg:prony} {\itshape Try Prony's Algorithm}. 

\vspace{-2ex}
\begin{algorithm}
\vspace*{1ex}
{\itshape Try Prony's algorithm}
\newline
\ruleaboveonelineheader
\label{alg:prony}
{\itshape Input:\/}
\begin{itemize}
\item[\ourtriang]
A position $r$ and sequence $(\eval_r,\ldots,\eval_{r+2B-1})$ in $\KK$ where
$\KK$ is a field of scalars. 
\item[\ourtriang]
An upper bound $D\in \ZZ_{>0}$. 
\item[\ourtriang]
$ \omega\in \KK\setminus\{0\} $ and has order $\geq 2D+1$.
\item[\ourtriang]
Algorithm~\ref{alg:intlog}: Integer Logarithm Algorithm that 
takes $D, \omega, \rho$ as input and
outputs:
\begin{itemize}
\item[\ourtriang]
either $\delta\in \ZZ$ with $|\delta|\leq D$ and
$\omega^{\delta}=\rho$,
\item[\ourtriang]
or FAIL.
\end{itemize}
\end{itemize}
{\itshape Output:\/}
\begin{itemize}
\item[\ourtriang]
Either 
a sparse Laurent 
polynomial of sparsity $t\leq B$ 
and has term degrees $\delta_j$ with $|\delta_j|\leq D$,
\item[\ourtriang] 
or FAIL.
\end{itemize}

{\setlength{\leftmargini}{\myleftmargini}

\begin{enumerate}[Step 1:]
\setlength{\itemindent}{\myitemindentstep}

\item
{\itshape Use Berlekamp/Massey algorithm to compute the
minimal linear generator of the sequence 
$ (\eval_{r},\ldots,\eval_{r+2B-1} ) $ and denote it
by $ \Lambda(z) $.
If $\Lambda(0) = 0$ return FAIL.
} 
	
\item {\itshape Compute all distinct roots $\in \KK$ of $ \Lambda(z)$,
denoted by $\rho_1,\ldots,\rho_t$.
If $t < \deg(\Lambda)$ then return FAIL.}
 

\item{\itshape For $j=1,\ldots, t$, use the {\upshape 
Algorithm \ref{alg:intlog}:} Integer 
Logarithm Algorithm to  compute
$\delta_j=\log_{\omega}\rho_j$. If the Integer Logarithm Algorithm
returns FAIL, then return FAIL.}

\item \label{step:delta_continue} 
{\itshape Compute the coefficients $ 
c_1,\ldots, c_t $ by solving the following transposed generalized 
Vandermonde system}
\begin{equation*}
	\begin{bmatrix}
		\rho_1^r & \rho_2^r & \cdots & \rho_t^r \\
		\rho_1^{r+1} & \rho_2^{r+1} & \cdots & \rho_t^{r+1}\\
		\vdots & \vdots & \vdots & \vdots \\
		\rho_1^{r+t-1} & \rho_2^{r+t-1} & \vdots & \rho_t^{r+t-1}
		\end{bmatrix}
		\begin{bmatrix}
			c_1\\
			c_2\\
			\vdots\\
			c_t
		\end{bmatrix}=
		\begin{bmatrix}
			\eval_r\\
			\eval_{r+1}\\
			\vdots\\
			\eval_{r+t-1}
		\end{bmatrix}.
	\end{equation*}

\item{\itshape Return the polynomial
$\sum_{j=1}^{t}c_jx^{\delta_j}$}.
\rulebelowtext{1.5ex}
\newline
\end{enumerate}
}
\end{algorithm}

\vspace{-2ex}

Now we give an algorithm for interpolating a 
black-box polynomial with sparsity bounded by $B$.
This algorithm can correct one error in $3B$ evaluations.
More specifically, if there is at most one error in the $3B$ 
evaluations of a univariate black-box polynomial 
$f(x)$ of sparsity $\leq B$, 
then the Algorithm~\ref{alg:power_1error} 
will compute a list of sparse 
interpolants containing $f(x)$.  
Moreover, $f(x)$ is not  
distinguishable from other interpolants (if there are any) 
in the list, because all interpolants returned by the 
algorithm satisfy the output conditions and $f(x)$ could be  
any one of them. In fact,   
\citeP[Theorem 3]{KaPe14} shows that for $\KK=\CC$, 
one needs $N\geq 2B(2E+1)$ points to  
guarantee a unique interpolant 
where $E\geq$ the number of errors.  
However, for $\KK=\RR$, by Descartes's rule of signs,  
\citeP[Corollary 2]{KaPe14} shows that if we probe  
$f(x)$ at $N\geq2B+2E$ distinct positive arguments 
with at most $E$ errors in the output evaluations, 
then $f(x)$ is the only interpolant in $\RR[x]$ which has  
sparsity $\leq B$ and has $\leq E$ errors in the  
$N$ evaluations. Therefore, in the  
Algorithm~\ref{alg:power_1error}, if $\KK=\RR$, $\omega>0$,  
$3B \geq 2B+2E=2B+2$, and the $3B$ evaluations contain at most 
one error, then $f(x)$ will be the only interpolant in the 
output.  
The Algorithm~\ref{alg:power_1error} will return  
FAIL if 
no sparse interpolants satisfy the output 
conditions.  

\vspace{-1ex} 
\begin{algorithm}
{\itshape A list-interpolation algorithm for power-basis sparse
polynomials with evaluations containing at most one error.}
\newline
\label{alg:power_1error}
\ruleabovetwolineheader
{\itshape Input:\/}
\begin{itemize}
\item[\ourtriang]
A black box representation of a polynomial $ f\in \KK[x,x^{-1}]$ 
where $ \KK $ is a field of scalars. 
The black box 
for $f$
returns the same (erroneous) output when 
probed multiple times at the same input.
\item[\ourtriang]
An upper bound $B$ on the sparsity of $f$.
\item[\ourtriang]
An upper bound $D\geq \max_j |\delta_j|$, 
where $\delta_j$ are term degrees of $f$.
\item[\ourtriang]
$ \omega\in \KK\setminus\{0\} $ satisfying: 
\begin{itemize}
\item[\ourtriang]
$\omega$ has order $\geq 2D+1$;
\item[\ourtriang]
$ \omega^{i_1}\neq \omega^{i_2} $ for all 
$ 1\leq i_1 < i_2\leq 3B $.
\end{itemize}
\item[\ourtriang]
An algorithm that computes all roots $\in \KK$ of a polynomial $\in \KK[x]$.
\end{itemize}
{\itshape Output:\/}
\begin{itemize}
\item[\ourtriang]
Either a list of sparse polynomials $\{ \fsupbr{1}, \ldots,
\fsupbr{M} \}$ with each $\fsupbr{k}$ 
$(1\leq k\leq M)$ satisfying:
\begin{itemize}
\item[\ourtriang]
$\fsupbr{k}$ has sparsity $ \leq B $ 
and has term degrees $\delta_j$ with $|\delta_j|\leq D$;
\item[\ourtriang]
$\fsupbr{k}$ is represented by its term
degrees and coefficients;
\item[\ourtriang]
there is at most one 
index $i\in \{1,2,\ldots, 3B\} $ such
that $\fsupbr{k}(\omega^i)\neq \eval_i$ where 
$\eval_i$ is the output of the black box probed 
at input $\omega^i$;
\item[\ourtriang]
$ f $ is contained in the list,
\end{itemize}
\item[\ourtriang]
or FAIL. 
\end{itemize}


{\setlength{\leftmargini}{\myleftmargini}

\begin{enumerate}[Step 1:]
\setlength{\itemindent}{\myitemindentstep}

\item
{\itshape For $ i=1,2,\ldots,3B $, get the
	 output $ \eval_i$ of the black box for $f$ at input  
	 $ \omega^i $.
	Let $ L $ be an empty list.}

\item \label{step:error_at_tail} 
{\itshape Use Algorithm~\ref{alg:prony} on the sequence
	$(\eval_{1},\eval_{2}, \ldots, \eval_{2B})$. 
	If the algorithm returns a sparse polynomial 
	$\bar{f}$ of sparsity $\leq B$ 
	and has term degrees $\delta_j$ with $|\delta_j|\leq D$, 
	and 
	there is at most one 
	index $i\in\{1,2,\ldots, 3B\}$ such that
	$\bar{f}(\omega^i)\neq \eval_i$, then
	add $\bar{f}$ to the list $L$.} 

If the error is in $(\eval_{2B+1},\eval_{2B+2}\ldots, 
\eval_{3B} )$, then the sequence 
$( \eval_1,\eval_2, \ldots,\eval_{2B} )$ is free of errors, so
Algorithm~\ref{alg:prony} 
in Step \ref{step:error_at_tail}  
will return $f$, and $f$ will be added to the list $L$.

\item \label{step:error_at_head} 
{\itshape Use Algorithm~\ref{alg:prony} on the sequence
	$(\eval_{B+1}$, 
$\eval_{B+2}, \ldots,$ $\eval_{3B})$. 
	If the algorithm returns a sparse polynomial 
	$\bar{f}$ of sparsity $\leq B$ 
	and has term degrees $\delta_j$ with $|\delta_j|\leq D$, 
	and 
	there is at most one 
	index $i\in\{1,2,\ldots, 3B\}$ such that
	$\bar{f}(\omega^i)\neq \eval_i$, then
	add $\bar{f}$ to the list $L$.} 

If the error is in $(\eval_1,\ldots, \eval_B )$,
then the sequence $(\eval_{B+1},\eval_{B+2},\ldots,\eval_{3B} )$ 
is free of errors, so Algorithm~\ref{alg:prony} in 
Step \ref{step:error_at_head}
will return $f$, and $f$ will be added into the list $L$.

\item \label{step:error_at_middle}
{\itshape For $ \ell = B+1,B+2, \ldots, 2B $,  }

\begin{enumerate}[{\theenumi}(a):]
\setlength{\itemindent}{\myitemindentonea}
	\item {\itshape substitute $ \eval_{\ell} $ by a symbol $
	\alpha $ in the matrix $ H_{\ell-B} $ 
    (see \eqref{eqn:Hankel}); use the fraction free
    Berlekamp/Massey algorithm \citeP{GKL02, KaYu08} 
    to compute the determinant of $ H_{\ell-B} $ 
    and denote it by
	$ \Delta_{\ell}(\alpha) $; }

Here $ \Delta_{\ell}(\alpha) $ is a univariate polynomial of the 
form $ (-1)^{B+1}\alpha^{B+1} + \tilde{\Delta}_{\ell}(\alpha)$ 
with $\deg (\tilde{\Delta}_{\ell}(\alpha)) < B+1 $;
	
	\item \label{step:comp_verify_xi} 
	{\itshape compute all solutions of the equation 
	$ \Delta_{\ell}(\alpha ) = 0 $ in $ \KK $; 
	denote the solution set as 
	$\{\xi_1, \ldots, \xi_b \} $ ;}

	\item {\itshape for $ k=1,\ldots,b $,} 
	\begin{enumerate}[{\theenumi}(\theenumii)i:]
\setlength{\itemindent}{\myitemindentoneaiii}
	\item {\itshape substitute $ \eval_{\ell} $ by $
		\xi_k $;}
	\item {\itshape use Berlekamp/Massey algorithm to compute the
		the minimal linear generator of the new sequence
		$(\eval_1,\eval_2,\ldots,\eval_{3B}) $ and
		denote it by $\Lambda(z)$;} 
	\item {\itshape if $\deg (\Lambda(z)) \leq B$, repeat Step 
		\ref{step:error_at_tail}.}
	\end{enumerate}
	
%
%

\end{enumerate}

If $\eval_{\ell}\neq f(\omega^{\ell})$ with 
$\ell \in \{B+1,B+2,\ldots, 2B \}$, then we substitute 
$\eval_{\ell}$ by a symbol $\alpha$ and compute the roots
$\{\xi_1, \ldots, \xi_b \} $ of
$\Delta_\ell(\alpha)$ in $\KK$. The correct value $f(\omega^\ell)$ is
in the set 
$\{\xi_1, \ldots, \xi_b \} $.
Thus for every root $\xi_k$
$(k=1,\ldots,b)$, we replace $\eval_{\ell}$ with $\xi_k$ and 
use Berlekamp/Massey algorithm to check if
the new sequence $(\eval_1,\eval_2,\ldots,\eval_{3B})$ is
generated by some polynomial of degree $\leq B$. If so, then
we apply Algorithm~\ref{alg:prony} 
on the updated sequence 
$(\eval_1,\eval_2,\ldots,\eval_{2B})$.
In the end, Step \ref{step:error_at_middle} will add $f$ into the list 
$L$ in case that $B+1\leq \ell\leq 2B$.

\item 
{\itshape If the list $L$ is empty, then return FAIL, otherwise return
the list $L$. }
\rulebelowtext{0.8ex}
\end{enumerate}
}
\end{algorithm}

\vspace{-2ex}

\begin{prop}\label{prop:power_outputbound_1error}
	The output list of Algorithm \ref{alg:power_1error} 
	contains $\leq B^2+B+2 $ polynomials.
\end{prop}
\begin{IEEEproof}
The Step~\ref{step:error_at_tail} 
in 
Algorithm~\ref{alg:power_1error} 
produces $ \leq 1 $ polynomial 
and so is Step~\ref{step:error_at_head}.  
In the Step~\ref{step:error_at_middle}    
of Algorithm~\ref{alg:power_1error},      
because $ \Delta_{\ell}(\alpha) $ 
has degree $ B+1 $, the equation $ 
\Delta_{\ell}(\alpha) = 0 $ has $ \leq B+1 $ solutions in $\KK$,
therefore this step produces $ \leq B(B+1) $ 
polynomials. Thus the output list of 
Algorithm~\ref{alg:power_1error} 
contains $ \leq 2+B(B+1) $ 
polynomials.
\end{IEEEproof}

\subsection{Correcting $2$ Errors}
\label{subsec:2_errors_power}
In this section, we give a list-interpolation algorithm to recover
$ f(x) $ (see \eqref{eqn:f_powerbasis}) from $ 4B $ evaluations 
that contain 2 errors. 
Recall that $ B $ is an upper bound on the sparsity of $ 
f(x) $ and $D$ is an upper bound on the absolute values of
the term degrees of $f(x)$. 
We will use Algorithm \ref{alg:power_1error}
as a subroutine. 

Let $ \omega\in \KK\setminus\{0\} $ such that:
(1) $\omega$ has order $\geq 2D+1$, and (2) 
$\omega^{i_1}\neq \omega^{i_2} $ for all $1\leq i_1 < i_2
\leq 4B$. 
For $ i=1,2,\ldots,4B $, let $ 
\eval_i $ be the output of the black box probed at input 
$\omega^i$. Let $\eval_{\ell_1}$ and $\eval_{\ell_2}$ be 
the 2 errors and $\ell_1 < \ell_2$. The problem can be 
covered by the following four cases: 
{\setlength{\leftmargini}{\myleftmargini}
\begin{enumerate}[{Case} 1:]
\setlength{\itemindent}{\myitemindentcase}
	
	\item \label{case:2errors_at_head_power}
	$1\leq \ell_1\leq B$;
	
	\item \label{case:2errors_at_tail_power}
	$3B+1 \leq \ell_2\leq 4B$;
    
    \item \label{case:2errors_at_eithermiddle}
    $B+1 \leq \ell_1<\ell_2\leq 2B$ or 
    $2B+1\leq \ell_1<\ell_2\leq 3B$;

	\item \label{case:2errors_at_middle_power}
	$B+1\leq \ell_1 \leq 2B$ and 
	$2B+1\leq \ell_2\leq 3B$.
\end{enumerate}
}

First, we try the Algorithm~\ref{alg:power_1error} on the
sequences $(\eval_1$, 
$\eval_2,\ldots,\eval_{3B} )$ and
$(\eval_{B+1},\eval_{B+2},\ldots,\eval_{4B})$, 
which can list interpolate
$f(x)$ if either 
Case~\ref{case:2errors_at_tail_power} 
or
Case~\ref{case:2errors_at_head_power} 
happens. 
Next, we use the Algorithm~\ref{alg:prony} 
on the sequences $(\eval_1,\ldots,\eval_{2B})$ and 
$(\eval_{2B+1},\ldots,\eval_{4B})$, which will return $f(x)$ 
if Case~\ref{case:2errors_at_eithermiddle} happens. 
For Case~\ref{case:2errors_at_middle_power}, 
we substitute the two
erroneous values $\eval_{\ell_1}$ and $\eval_{\ell_2}$ by
two symbols
$\alpha_1$ and $\alpha_2$ 
respectively. Then the pair
of correct values $(f(\omega^{\ell_1}),f(\omega^{\ell_2}))$
is a solution of the following Pham system (see 
Lemma~\ref{lem:2det_neq_0} and 
Lemma~\ref{lem:Pham_sys}): 
\begin{equation}\label{eqn:Pham_sys}
	\det(\sHankel_{\ell_1-B}) = 0,\ 
	\det(\sHankel_{\ell_2-B}) = 0,
\end{equation} 
where $\sHankel_{\ell_1-B}$ and $\sHankel_{\ell_2-B}$ are 
Hankel matrices defined as \eqref{eqn:Hankel}.
As the 
Pham system 
\eqref{eqn:Pham_sys} is zero-dimensional 
(see Lemma~\ref{lem:Pham_sys}), 
we compute the solution set 
$\{(\xi_{1,1},\xi_{2,1}),\ldots,(\xi_{1,b},\xi_{2,b}) \} $ 
of \eqref{eqn:Pham_sys}.
Then, for $k=1,\ldots,b$, we substitute
$(\eval_{\ell_1}, \eval_{\ell_2})$ by 
$(\xi_{1,k},\xi_{2,k})$
and apply Algorithm~\ref{alg:prony} on the updated sequence
$(\eval_1,\eval_2, \ldots, \eval_{2B})$;
this results in a list of candidates for $f$ if 
Case~\ref{case:2errors_at_middle_power} 
happens.

The following Lemma shows that the determinants arising 
in~\eqref{eqn:Pham_sys} have the Pham property, using 
diagonals in place of anti-diagonals. 

\begin{lemma}\label{lem:2det_neq_0} 
Let $ A $ be an $ n\times n $ matrix with the following 
properties: 
{\setlength{\leftmargini}{\wd\conditionone} 
\begin{enumerate}[\upshape(1)]
	\item for $ i=1,\ldots,n $, 
    $ A[i,i] = \alpha_1 $;
	\item for some fixed $k\in \{1,\ldots,n-1\}  $ and for 
	$ i=1,\ldots, n-k $, 
    $ A[i,i+k]=\alpha_2 $;
	\item all other entries of 
the matrix $A$ are elements 
in the field of scalars $\KK$. 
\end{enumerate} 
}
Then 
$ \text{det}(A)=\alpha_1^n + Q(\alpha_1,\alpha_2) $ 
where $ Q(\alpha_1,\alpha_2) $ is a polynomial
of total degree $ \leq n-1 $.  
\end{lemma}

\begin{IEEEproof} The matrix $ A $ is of the form:
	\begin{equation*} 
	A = \begin{bmatrix} 
	\alpha_1 & \cdots & \alpha_2 &        &   *    \\ 
	\quad  & \ddots &\ddots & \ddots &  \quad  \\ 
		   &        &\ddots & \ddots & \alpha_2 \\ 
		   &    *   &       & \ddots & \vdots \\ 
		   &        &       &        &\alpha_1 \\
        \end{bmatrix}.  
    \end{equation*} 

    We prove by induction on $ n $. It is trivial if $n=1$. 
    Assume that the conclusion holds for $ n-1 $. By
	minor expansion on the first column of $ A $, we
	have 
	\begin{equation*} 
	\det(A) = \alpha_1(\alpha_1^{n-1}
	+ Q_1(\alpha_1,\alpha_2) ) + Q_2(\alpha_1,\alpha_2)
	\end{equation*} 
	where $ Q_2(\alpha_1,\alpha_2) $ has total degree $\leq n-1 $. 
    By induction hypothesis, $ Q_1(\alpha_1,\alpha_2) $ has 
	total degree $\leq n-2 $. Let $Q = \alpha_1\cdot Q_1 + Q_2$. 
    The proof is complete.
\end{IEEEproof}

\begin{lemma}\label{lem:Pham_sys} 
	The Pham system
	\begin{equation} 
	\begin{split} 
	& \alpha_1^{n_1} + Q_1(\alpha_1, \alpha_2)=0, \ \deg (Q_1)\leq n_1 - 1 
	\\ 
	&\alpha_2^{n_2} + Q_2(\alpha_1, \alpha_2)=0,\ \deg
	(Q_2) \leq n_2 - 1
	\end{split} 
	\end{equation} 
	has at most 
$n_1 \times n_2$ 
solutions, where $ Q_1 $ and $ 
	Q_2 $ are two polynomials in $ \KK[\alpha_1, \alpha_2] $.
\end{lemma}

\begin{IEEEproof} 
	See e.g. \citeP[Chapter 5, Section 3, Theorem 
	6]{CLO2015ideals}.  
\end{IEEEproof}

\begin{example}
Let $B=3$. With $4B=12$ evaluations
$\eval_1,\eval_2,\ldots,\eval_{12}$ obtained from the black box
for $f$ at inputs $\omega, \omega^2,\ldots, \omega^{12}$,
we have the following $9\times 4$ matrix:
\begin{equation*}
\sHankel =
\begin{bmatrix}
\eval_1 & \eval_2 & \eval_3 & \eval_4 \\
\eval_2 & \eval_3 & \eval_4 & \eval_5 \\
\eval_3 & \eval_4 & \eval_5 & \eval_6 \\
\eval_4 & \eval_5 & \eval_6 & \eval_7 \\
\eval_5 & \eval_6 & \eval_7 & \eval_8 \\
\eval_6 & \eval_7 & \eval_8 & \eval_9 \\
\eval_7 & \eval_8 & \eval_9 & \eval_{10} \\
\eval_8 & \eval_9 & \eval_{10} & \eval_{11} \\
\eval_9 & \eval_{10} & \eval_{11} & \eval_{12}\\ 
\end{bmatrix}\in \KK^{9\times 4}
\end{equation*}

Suppose there are two errors $\eval_{\ell_1}, \eval_{\ell_2}
(\ell_1 < \ell_2)$ in the evaluations.
If $\ell_1\in \{1,2,3\}$, then the Algorithm
\ref{alg:power_1error} can recover $f(x)$ from the last $3B$
evaluations $(\eval_{4}, \eval_5, \ldots, \eval_{12})$.
Similarly, $f(x)$ can also be recovered from
$(\eval_1,\eval_2,\ldots,\eval_9)$ by the Algorithm~\ref{alg:power_1error} 
if $ \ell_2\in \{10, 11, 12\}$.
Next, if $\ell_1,\ell_2\in \{4,5,6\}$     
or $\ell_1,\ell_2\in \{7,8,9\}$, then the        
Algorithm~\ref{alg:prony} can recover $f(x)$   
from $(\eval_7,\ldots,\eval_{12} )$ or 
$(\eval_1,\ldots,\eval_6)$.        

It remains 
to consider the case that
$\ell_1\in \{4,5,6\} $ and
$\ell_2\in \{7,8,9\}$. We substitute $\eval_{\ell_1}$,
$\eval_{\ell_2} $ by 
$ \alpha_1 $, $\alpha_2$
respectively. Then
the determinants of the matrices $\sHankel_{\ell_1-3}$ and
$\sHankel_{\ell_2-3}$ can be written as:
\begin{equation}\label{example:Pham_system}
\begin{split}
&\det(\sHankel_{\ell_1-3} ) = -\alpha_1^4 +
Q_1(\alpha_1,\alpha_2),\; \deg Q_1\leq 3\\
& \det(\sHankel_{\ell_2-3} ) = -\alpha_2^4 +
Q_2(\alpha_1, \alpha_2),\; \deg Q_2 \leq 3
\end{split}
\end{equation}
where $\sHankel_{\ell_1-3} $, $\sHankel_{\ell_2-3} $ are Hankel
matrices defined as \eqref{eqn:Hankel} and
where $Q_1$ and $Q_2$ are bivariate polynomials in
$\alpha_1$ and $\alpha_2$.
We compute the roots
$(\xi_{1,k},\xi_{2,k})_{k\geq 1}$
of the system
\eqref{example:Pham_system} in $\KK$ and 
the pair of correct 
values $(f(\omega^{\ell_1}), f(\omega^{\ell_2}))$ is one of the
roots. For each root 
$(\xi_{1,k},\xi_{2,k})$,
we substitute
$\eval_{\ell_1}$, $\eval_{\ell_2}$ by 
$\xi_{1,k}$, $\xi_{2,k}$
respectively, and check if the matrix $\sHankel$ has
rank $B=3$. If so, then run Algorithm~\ref{alg:prony} on the
updated sequence $(\eval_1,\eval_2,\ldots,\eval_6)$. In the end,
we obtain a list of sparse polynomials that contains
$f(x)$.
\end{example}

We summarize the process of correcting $2$ errors 
in 
Algorithm~\ref{alg:power_2errors} below. 
If there are at most $2$ errors in the $4B$ 
evaluations of a univariate black-box polynomial 
$f(x)$ of sparsity $\leq B$, 
then the Algorithm~\ref{alg:power_2errors} 
will compute a list of sparse 
interpolants containing $f(x)$.  
Again, $f(x)$ is not  
distinguishable from other interpolants (if there are any) 
in the list, because all interpolants returned by the 
algorithm satisfy the output conditions and $f(x)$ could be  
any one of them.   
Nevertheless, in the Algorithm~\ref{alg:power_2errors},  
if $\KK=\RR$, $\omega>0$,  
$4B \geq 2B+2E=2B+4$, and the $4B$ evaluations contain  
at most 
$2$ errors, then $f(x)$ will be the only  
interpolant in the 
output.  
The Algorithm~\ref{alg:power_2errors} will return  
FAIL if no 
sparse interpolants satisfy the output 
conditions.  

\vspace{-1ex} 
\begin{algorithm}
{\itshape A list-interpolation algorithm for power-basis
sparse polynomial with evaluations containing at most $2$ errors.}
\newline
\ruleabovetwolineheader 
\label{alg:power_2errors}
{\itshape Input:\/}
\begin{itemize}
\item[\ourtriang]
A black box representation of a polynomial $ f\in \KK[x,x^{-1}]$ 
where $ \KK $ is a field 
of scalars. 
The black box for $f$
returns the same (erroneous) output when 
probed multiple times at the same input.

\item[\ourtriang]
An upper bound $B$ on the sparsity of $f$.

\item[\ourtriang]
An upper bound $D\geq \max_j |\delta_j|$, 
where $\delta_j$ are term degrees of $f$.

\item[\ourtriang]
$ \omega\in \KK\setminus\{0\} $ satisfying:
\begin{itemize}
\item[\ourtriang] 
$\omega$ has order $\geq 2D+1$;
\item[\ourtriang]
$ \omega^{i_1}\neq \omega^{i_2} $ for all
$ 1\leq i_1 < i_2\leq 4B $.
\end{itemize}

\item[\ourtriang]
An algorithm to compute all roots $\in \KK$ of polynomials in $\KK[x]$.
\end{itemize}
{\itshape Output:\/}
\begin{itemize}
\item[\ourtriang]
Either a list of sparse polynomials 
$\{ \fsupbr{1}, \ldots,\fsupbr{M} \}$ with 
each $\fsupbr{k}$ $(1\leq k\leq M)$ satisfying:

\begin{itemize}
\item[\ourtriang]
$\fsupbr{k}$ has sparsity $ \leq B $ 
and has term degrees $\delta_j$ with $|\delta_j|\leq D$;

\item[\ourtriang]
$\fsupbr{k}$ is represented by its term
degrees and coefficients; 

\item[\ourtriang]
there are at most $2$ 
indices $i_1, i_2\in \{1,2,\ldots, 4B\} 
$ such that $\fsupbr{k}(\omega^{i_1})\neq \eval_{i_1}$ 
and $\fsupbr{k}(\omega^{i_2})\neq \eval_{i_2}$ where 
$\eval_{i_1}$ and $ \eval_{i_2} $ are the outputs of the 
black box probed at inputs $\omega^{i_1}$ and $ \omega^{i_2} $ 
respectively;

\item[\ourtriang]
$f$ is contained in the list,
\end{itemize}

\item[\ourtriang]
or FAIL. 
\end{itemize}


{\setlength{\leftmargini}{\myleftmargini}

\begin{enumerate}[Step 1:]
\setlength{\itemindent}{\myitemindentstep}

\item
{\itshape For $ i=1,2,\ldots,4B $, get the output $ 
\eval_i $ of the black box for $f$ at input $ \omega^i $.}

\item \label{step:1error_at_tailorhead}
{\itshape Take $ (\eval_1,\eval_2,\ldots, \eval_{3B} )$ and 
$( \eval_{B+1},\eval_{B+2},\ldots, \eval_{4B} )$  as the 
evaluations at the first step of Algorithm 
\ref{alg:power_1error} and get two lists 
$L_1$ and $L_2$, 
where $L_1$ and $L_2$ are either the output 
lists of Algorithm~\ref{alg:power_1error} or empty lists if 
Algorithm~\ref{alg:power_1error} returns FAIL. 
Let $ L $ be the union of $ L_1 $ and $ L_2 $.}

If either $ (\eval_1,\eval_2,\ldots, 
\eval_{3B} )$ or $( \eval_{B+1},\eval_{B+2},\ldots, \eval_{4B} )$
contains $\leq 1$ error, 
the Algorithm~\ref{alg:power_1error} can compute 
a list of sparse polynomials containing $ f(x) $.

\item \label{step:2errors_at_tailorhead} 
{\itshape Use Algorithm~\ref{alg:prony} on the sequences  
$(\eval_1, \eval_2, \ldots,\eval_{2B})$ and  
$(\eval_{2B+1}, \eval_{2B+2}, \eval_{4B})$. If 
Algorithm~\ref{alg:prony} returns a sparse polynomial  
$\bar{f}$ of sparsity $\leq B$ and has 
term degrees $\delta_j$ with $|\delta_j|\leq D$, then add 
$\bar{f}$ into the list $L$. 
}

If either $(\eval_1, \eval_2, \ldots,\eval_{2B})$ or  
$(\eval_{2B+1}, \eval_{2B+2}, \eval_{4B})$  
is error-free, the Algorithm
~\ref{alg:prony} will return  
$f(x)$. 

\item
{\itshape For every polynomial $ \bar{f} $ in the list $ L 
$, if 
there are at least $3$ 
indices $i\in \{1,2,\ldots, 
4B\} $ such that $ \bar{f}(\omega^i)\neq\eval_i  $ then 
delete $ \bar{f} $ from $ L $.}

\item \label{step:2errors_in_middle}
{\itshape For $ \ell_1=B+1,\ldots, 2B $ and 
$\ell_2=2B+1,\ldots,3B $,}

\begin{enumerate}[\theenumi (a):]
\setlength{\itemindent}{\myitemindentonea}
	
	\item {\itshape substitute $ \eval_{\ell_1} $ by $ 
	\alpha_1 $ and $ \eval_{\ell_2} $ by $ \alpha_2 $ in the 
	Hankel matrices $ \sHankel_{\ell_1-B} $ and $ 
	\sHankel_{\ell_2-B} $ (see \eqref{eqn:Hankel}); 
    let $ \Delta_{\ell_1}(\alpha_1,\alpha_2)= 
	\det(\sHankel_{\ell_1-B}) $ 
	and $ \Delta_{\ell_2}(\alpha_1,\alpha_2)= 
	\det(\sHankel_{\ell_2-B}) $.}
    
    Here, we also use the fraction free Berlekamp/Massey
    algorithm \citeP{GKL02,KaYu08} to compute the determinants
    of $ \sHankel_{\ell_1-B} $ and $ \sHankel_{\ell_2-B} $.

	\item {\itshape compute all solutions of the Pham system $ 
	\{ \Delta_{\ell_1}(\alpha_1,\alpha_2) = 0, 
	\Delta_{\ell_2}(\alpha_1,\alpha_2) = 0 \}$  in $ \KK^2 $; 
	denote the solution set as 
    $\{(\xi_{1,1},\xi_{2,1}),\ldots, (\xi_{1,b},\xi_{2,b})\} $.
}
\newline
One may use a Sylvester resultant algorithm and the root finder
in $\KK[x]$ to accomplish this task in polynomial time.

	\item \label{step:verify_2xi}
	{\itshape for $ k=1,\ldots,b $,}
	\begin{enumerate}[\theenumi(\theenumii)i:]
\setlength{\itemindent}{\myitemindentoneaiii}
		
\item {\itshape substitute $ \eval_{\ell_1}$
by $ \xi_{1,k}$ and $ \eval_{\ell_2} $ by 
$ \xi_{2,k} $;}

\item {\itshape use Berlekamp/Massey algorithm to compute the
the minimal linear generator of the new sequence
$(\eval_1,\eval_2,\ldots,\eval_{4B}) $ and
denote it by $\Lambda(z)$; }

\item {\itshape if $\deg (\Lambda(z)) \leq B$,
use Algorithm~\ref{alg:prony} on the updated
sequence $ (\eval_1,\eval_2, \ldots, \eval_{2B}) $; if 
Algorithm~\ref{alg:prony} returns a sparse polynomial 
$\bar{f}$ of sparsity $\leq B$ 
and has term degrees $\delta_j$ with $|\delta_j|\leq D$, 
and 
there are at most $2$ 
indices $i_1,i_2 \in \{1,2, \ldots, 
4B\} $ such that $ \bar{f}(\omega^{i_1})\neq 
\eval_{i_1} $ and $ \bar{f}(\omega^{i_2})\neq 
\eval_{i_2} $, then add $ \bar{f} $ into the list 
$ L $;}
		
\end{enumerate}
	
\end{enumerate}

If the two errors are $ \eval_{\ell_1} $ and $ 
\eval_{\ell_2} $ with $\ell_1\in \{B+1,\ldots,2B\}$ and $
\ell_2\in \{2B+1,\ldots,3B\} $, we substitute 
$ \eval_{\ell_1} $ and  $ \eval_{\ell_2} $ by two 
symbols $ \alpha_1 $ and $ \alpha_2 $ respectively. As the 
pair of correct values 
$ (f(\omega^{\ell_1}), f(\omega^{\ell_2}) ) $ 
is a solution of the system 
$ \{ \Delta_{\ell_1}(\alpha_1,\alpha_2) = 0, 
\Delta_{\ell_2}(\alpha_1,\alpha_2) = 0 \} $, Step 
\ref{step:2errors_in_middle} will add $ f $ into the 
list $ L $.

\item 
{\itshape If the list $L$ is empty, then return FAIL, otherwise return
the list $L$.}
\rulebelowtext{0.8ex}
\end{enumerate}
}
\end{algorithm}

\vspace{-2ex}

\begin{prop}\label{prop:power_outputbound_2errors}
The output list of Algorithm~\ref{alg:power_2errors} 
contains $\leq B^4+2B^3+3B^2+2B+6 $ 
polynomials.
\end{prop}
\begin{IEEEproof}
In Algorithm~\ref{alg:power_2errors}, 
only 
Step~\ref{step:1error_at_tailorhead}, 
Step~\ref{step:2errors_at_tailorhead}, 
and Step~\ref{step:2errors_in_middle} produce new polynomials. 
By Proposition~\ref{prop:power_outputbound_1error}, 
both the lists $ L_1 $ and $ L_2 $ obtained at 
Step~\ref{step:1error_at_tailorhead} 
contain $ \leq B^2+B+2 $ polynomials. 
Step~\ref{step:2errors_at_tailorhead} produces $\leq 2$ 
polynomials. 
For Step~\ref{step:2errors_in_middle} of 
Algorithm~\ref{alg:power_2errors}, the Pham system $ 
\{ \Delta_{\ell_1}(\alpha,\beta) = 0, 
\Delta_{\ell_2}(\alpha,\beta) = 0 \}$ has $ \leq 
(B+1)^2 $ solutions, so this step produces $ \leq 
B^2(B+1)^2 $ polynomials. Therefore the output list 
contains  $\leq B^2(B+1)^2+2(B^2+B+2 )+2 $ 
polynomials.
\end{IEEEproof}

\subsection{Correcting $ E $ Errors} 
\label{subsec:E_errors_power}

Recall that $f(x)$ is a sparse univariate polynomial of the
form $\sum_{j=1}^{t} c_j x^{\delta_j}$ (see
\eqref{eqn:f_powerbasis}) with $t\leq B$ and $\forall j, |\delta_j|\leq D$. 
We show how to list interpolate $f(x)$ from $\numofevals$ 
evaluations containing $\leq E $ errors, where
\begin{equation}\label{eqn:number_of_evaluations}
\numofevals = \left\lfloor\frac{4}{3}E + 2\right\rfloor B.
\end{equation}
Let $\numofblocks = \left\lfloor E/3 \right\rfloor$. 
Choose 
$\omega_{1},\ldots,\omega_{\numofblocks},\omega_{\numofblocks
 + 1}  \in \KK\setminus\{0\}$ such that: 
{\setlength{\leftmargini}{\wd\conditionone} 
\begin{enumerate}[(1)]
\item $\omega_{\sigma}$ 
has order $\geq 2D+1$ for all $1\leq \sigma\leq
\numofblocks + 1$, and 
\item $\omega_{\sigma_1}^{i_1} \neq 
\omega_{\sigma_2}^{i_2}$ for any $1\leq \sigma_1 < \sigma_2\leq 
\numofblocks + 1 $ and $1\leq i_1 < i_2\leq 4B$.
\end{enumerate}
}
Let $\eval_{\sigma,i}$ denote the output of the black box 
at input $\omega_{\sigma}^{i}$ 
for $\sigma=1,\ldots,\numofblocks+1$ and $i=1,\ldots,4B$. 

If $E \bmod 3 = 0$ then $\numofevals = (E/3) 
4B + 2B$. The problem is reduced to 
one of the following 
situations: (1) the last block $ ( \eval_{\numofblocks+1,1},
\eval_{\numofblocks+1,2}, \ldots,  \eval_{\numofblocks+1,2B} ) 
$ of length $2B$ is free of error, or (2) there is some block $( 
\eval_{\sigma,1},\eval_{\sigma,2}, \ldots, \eval_{\sigma,4B} )$ 
with $1\leq \sigma\leq  E/3 $ which contains $\leq 2$ errors.  
These two situations can be 
handled 
by 
Algorithm~\ref{alg:prony} and Algorithm~\ref{alg:power_2errors}, respectively. 

If $E \bmod 3 = 1 $ then $\numofevals =
4B \numofblocks  + 3B$. The problem is reduced to 
one of the 
following situations: (1) the last block $ ( 
\eval_{\numofblocks+1,1},\eval_{\numofblocks+1,2},\ldots,  
\eval_{\numofblocks+1,3B} ) $ of length $ 3B $ has $\leq 
1$ error, or (2) 
there is some block $(
\eval_{\sigma,1},\eval_{\sigma,2},\ldots, \eval_{\sigma,4B} )$ 
with $1\leq \sigma\leq  \numofblocks $ which contains $\leq 2$ errors. 
Therefore by applying the 
Algorithm \ref{alg:power_1error} on $ (
\eval_{\numofblocks+1,1},\eval_{\numofblocks+1,2},\ldots,  
\eval_{\numofblocks+1,3B} ) $ and the Algorithm 
\ref{alg:power_2errors} on $( 
\eval_{\sigma,1},\eval_{\sigma,2},\ldots, \eval_{\sigma,4B} )$, we can list 
interpolate $f(x)$.
%

If $E\bmod 3 = 2$ then $E= 
3\,\numofblocks + 2$ and $\numofevals =( \numofblocks+1 
) 4B $. So there is some 
$\sigma\in \{1,\ldots,\numofblocks+1\}$ 
such that the block 
$(\eval_{\sigma,1},\eval_{\sigma,2},\ldots, \eval_{\sigma,4B})$ 
of length $4B$ contains $\leq 2$ errors, and we can use
the Algorithm \ref{alg:power_2errors} on this block to list 
interpolate $f(x)$.

\begin{remark}\label{rmk:power_outputbound_Eerrors}
We apply the Algorithm~\ref{alg:power_2errors} 
on every block $ (\eval_{\sigma,1}, \eval_{\sigma,2}, 
\ldots, \eval_{\sigma,4B}) $ for all 
$\sigma\in \{1,\ldots, \left\lfloor 
E/3 \right\rfloor \}  $, 
which will result in $\leq \left\lfloor 
E/3 \right\rfloor (B^4+2B^3+3B^2+2B+6 ) $  
polynomials according to 
Proposition~\ref{prop:power_outputbound_2errors}. 
The length of the last block depends on the value of $ 
E $, and we have the following different upper bounds on 
the number of resulting polynomials:
{\setlength{\leftmargini}{\wd\conditionone} 
\begin{enumerate}[(1)]
\item $ (E/3) (B^4+2B^3+3B^2+2B+6 )+1$, if $ 
E\bmod 3 = 0 $;
\item $ \left\lfloor 
E/3 \right\rfloor (B^4+2B^3+3B^2+2B+6 ) + 
B^2+B+2 $, if $ E\bmod 3=1 $ (see 
Proposition~\ref{prop:power_outputbound_1error}); 
\item $\left(\left\lfloor E/3 \right\rfloor 
+ 1\right)(B^4+2B^3+3B^2+2B+6 )$, if $ E\bmod 3=2 $.
\end{enumerate}
}
\end{remark}

By Descartes' rule of signs (see e.g. 
\citeP[Proposition~1.2.14]{BCR1998}), 
the approach for correcting $E$ errors
will produce a single polynomial if $\KK=\RR$, $N\geq 2B+2E$ and
$\omega_{\sigma}>0, \forall \sigma$. 
However, if $N < 2B+2E$ then there can be $\geq
2$ valid sparse interpolants. We give an example to illustrate this.


\begin{example}\label{example:power_list}
Choose $\omega > 0$. Let $B$ be an upper bound on the
sparsity of $f$ and $E$ be an upper bound on the number 
of errors in the evaluations. Let
\begin{equation*}
h = \prod_{i=0}^{2B-2}(x-\omega^i),
\end{equation*}
and $\fsupbr{1}$ be the sum of odd degree terms of $h$ 
and $\fsupbr{2}$ be the negative of the sum of even degree terms of $h$. 
Clearly, we have $h= \fsupbr{1}-\fsupbr{2} $ and 
$\fsupbr{1}(\omega^i) = \fsupbr{2}(\omega^i)$ for $i = 0,1, \ldots, 2B-2$.
Moreover, both $\fsupbr{1}$ and $\fsupbr{2}$ have sparsity $\leq B$ as
$\deg(h) =  2B -1 $.  
Consider a  sequence $\eval$ consisting of the following $2B+2E-1$ values:
\begin{equation}\label{eq:overreal_evals}
\begin{tabular}{%
@{\hspace{0.2em}}l@{\hspace{0.2em}}
@{\hspace{0.2em}}l@{\hspace{0.2em}}
@{\hspace{0.2em}}l@{\hspace{0.2em}}
@{\hspace{0.5em}}l@{\hspace{0.5em}}
@{}l@{}
}
$ a^{(1)}$
& $=$ 
& $\left( \fsupbr{1}(\omega^0) \right.$,  
& $\ldots,$ 
& $\left. \fsupbr{1}(\omega^{2B-2}) \right),$\\ 
$ a^{(2)}$
& $=$ 
& $\left( \fsupbr{1}(\omega^{2B-1}), \right.$ 
& $\ldots,$
& $\left. \fsupbr{1}(\omega^{2B+E-2}) \right),$\\
$ a^{(3)}$
& $=$ 
& $\left( \fsupbr{2}(\omega^{2B+E-1}), \right.$ 
& $\ldots,$ 
& $\left. \fsupbr{2}(\omega^{2B+2E-2}) \right),$\\
\end{tabular}
\end{equation}
that is, $\eval=(a^{(1)}, a^{(2)}, a^{(3)})$.
If all the errors are in $a^{(3)}$ 
then $\fsupbr{1}$ is a valid
interpolant. Alternatively, if all the errors are in 
$a^{(2)}$ 
then $\fsupbr{2}$ is a valid interpolant. Therefore, from these
$2B+2E-1$ values, we have at least $2$ valid interpolants.

We remark that one of the valid interpolants, 
$ \fsupbr{1}$ and $\fsupbr{2}$, must have $B$ terms since otherwise
uniqueness is guaranteed by Descartes's rule of signs. In this
example, both $ \fsupbr{1}$ and $\fsupbr{2}$ have $B$ terms because 
the polynomial $h$ has $2B$ terms. 
Indeed, $\deg(h)=2B-1$ implies that $h$ has $\leq
2B$ terms, and by Descartes' rule of signs, $h$ has $\geq 2B$ terms
because it has $2B-1$ positive real roots. Therefore $h$ is a dense
polynomial. However, 
with the following substitutions
\[
x = y^k,\; \omega = \bar{\omega}^k \text{ for some } k\gg 1,
\]
we have again a counter example where
$h$, $\fsupbr{1}$, and $\fsupbr{2}$ are sparse 
with respect to the new variable $y$.
\end{example}

\section{Sparse Interpolation in Chebyshev Basis with 
Error Correction}

\subsection{Correcting One Error}
\label{subsec:cheb_1error}

Let $ \KK $ be a field of 
scalars with characteristic $\ne 2$. 
Let $ f(x)\in \KK[x] $ be a polynomial represented by a black 
box. Assume that $f(x)$ is a sparse polynomial in
\chebT basis of the form:
\begin{multline*}
f(x) = \sum_{j=1}^t c_j T_{\delta_j}(x)\in\KK[x],\\
0 \le \delta_1 < \delta_2 < \cdots < \delta_t = \deg(f),
\forall j, 1\le j\le t\colon c_j \ne 0,
\end{multline*}
where $T_{\delta_j}(x)$
$(j=1,\ldots,t)$ are Chebyshev polynomials of the First
kind of degree $\delta_j$. 
We are given an upper bound $B\geq t$, and we want 
to recover term degrees $\delta_j$ and the         
coefficients $c_j$ from $3B$ evaluations of $f(x)$ 
where the evaluations contain at most one error.   
Using the formula
$T_{n}(\frac{x+x^{-1}}{2})=\frac{x^n+x^{-n}}{2}$ 
for all $ n\in \ZZ_{\geq 0} $, \citeP[Sec.~4]{ArKa15} 
transforms $f(x)$ into a sparse Laurent polynomial:
\begin{equation}\label{eqn:sparse_Laurent} 
	g(y)\defequal f(\frac{y+y^{-1}}{2}) =  \sum_{j=1}^{t} 
	\frac{c_j}{2}(y^{\delta_j}+y^{-\delta_j}). 
\end{equation}
Therefore the problem is reduced to recover the term degrees
and coefficients of the polynomial $g(y)$. 
Let $ \omega\in \KK $ such that 
$\omega$ has order $\geq 4D+1$. 

For $ i=1,2,\ldots,3B $, let $\eval_{2i-1}$ be the output of
the black box probed at input $
\gamma_{2i-1}=(\omega^{2i-1}+\omega^{-(2i-1)})/2 $. Note that
$g(\omega^i)=g(\omega^{-i})$ for any integer $i$.
For odd integers $ r\in \{2k-1\mid k=1,\ldots, B \} $, let 
$\sfoldmatrix_r\in \KK^{(B+1)\times(B+1)}$ be the following 
Hankel$+$Toeplitz matrix: 
\begin{equation}\label{eqn:fold_matrix}
	\sfoldmatrix_r = 
	\underbrace{\left[ \eval_{|r+2(i+j)|}
    \right]_{i,j=0}^B}_{\text{Hankel matrix}} + 
    \underbrace{\left[ \eval_{|r+2(i-j)|}
    \right]_{i,j=0}^B}_{\text{Teoplitz matrix}}.
\end{equation}
If all the values involved in the matrix $\sfoldmatrix_r$ 
are correct, then $\det(\sfoldmatrix_r) = 0$ 
\citeP[Lemma~3.1]{ArKa15}. 


If the $2B$ evaluations $\{\eval_{2i-1}\}_{i=1}^{2B}$
are free of errors, then one can use Prony's algorithm 
to recover $g(y)$ (and $f(x)$) from the following sequence
\citeP[Lemma~1]{KaPe14}:
\begin{equation}\label{seq:values_cheb}
\eval_{-4B+1}, \eval_{-4B+3}, \ldots, \eval_{-1},
\eval_{1}, \ldots,\eval_{4B-3}, \eval_{4B-1}.
\end{equation}
Now we show how to list interpolate $f(x)$ from $3B$ evaluations
$\{\eval_{2i-1}\}_{i=1}^{3B}$ containing $\leq 1$ error.

Assume that $\eval_{2\ell-1}$ is the error, that is,
$\eval_{2\ell-1}\neq f(\gamma_{2\ell-1})=g(\omega^{2\ell-1})$. 
The problem can
be reduced to three cases:
{\setlength{\leftmargini}{\myleftmargini}
\begin{enumerate}[{Case} 1:]
\setlength{\itemindent}{\myitemindentcase}
	
	\item\label{case:1error_at_head_cheby}
	$1\leq \ell\leq B$;
	
	\item\label{case:1error_at_middle_cheby}
	$B+1\leq \ell\leq 2B$;
	
	\item\label{case:1error_at_tail_cheby}
	$2B+1\leq \ell\leq 3B$.

\end{enumerate} 
}
For Case~\ref{case:1error_at_tail_cheby}, we can recover $f(x)$ from 
the sequence $(\eval_{2i-1} )_{i=-(2B-1)}^{2B}$.  
For the Case~\ref{case:1error_at_head_cheby} and 
Case~\ref{case:1error_at_middle_cheby}, we substitute 
$\eval_{2\ell-1}$ by a symbol $\alpha$. Let
\begin{equation*}
	\Delta_{2\ell-1}(\alpha) = 
\left\{
	\begin{tabular}{ll}
		$\det(\sfoldmatrix_{2\ell-1})$,    & if $1\leq \ell\leq B$,
                \\
		$\det(\sfoldmatrix_{2(\ell-B)-1})$,& if $B+1\leq \ell\leq 2B$,
	\end{tabular}
\right.
\end{equation*}
where $\sfoldmatrix_{2\ell-1} $ and $\sfoldmatrix_{2(\ell-B)-1} $ are
defined as in \eqref{eqn:fold_matrix} and $\Delta_{2\ell-1}(\alpha)$ 
is a univariate polynomial of degree
$B+1$ in $\alpha$ (see Lemma \ref{lem:foldmatrix_detneq0}).
By \citeP[Lemma~3.1]{ArKa15}, 
the correct value  
$f(\gamma_{2\ell-1})$ is a solution of the equation
$\Delta_{2\ell-1}(\alpha) = 0$. So we compute all solutions
$\{\xi_{1},\ldots,\xi_{b} \}$ of
$\Delta_{2\ell-1}(\alpha) = 0$ in $\KK$. For each solution
$\xi_k (1\leq k\leq b)$ we replace $\eval_{2\ell-1}$ by
$\xi_k$ and try Prony's algorithm 
on the updated sequence $(\eval_{2i-1} )_{i=-(2B-1)}^{2B}$.
In the end, we will get a list of polynomials 
containing $f(x)$. 

\begin{lemma}\label{lem:foldmatrix_detneq0}
	Let 
	$ r\in \{2k-1\mid k=1,\ldots, B \} $ and 
    $ \sfoldmatrix_r = \left[ \eval_{|r+2(i+j)|} + \eval_{|r+2(i-j)|}  
	\right]_{i,j=0}^B $.
    If $ \eval_r $ or $ \eval_{r+2B} $ is substituted by a 
	symbol $ \alpha $ in $ 
	\sfoldmatrix_r  $, then the determinant of 
	$ \sfoldmatrix_r $ is a univariate 
	polynomial of degree $ B+1 $ in $ \alpha $.
\end{lemma}
\begin{IEEEproof} 
First, we show that if $ \eval_{r+2B} $ is substituted by $ \alpha $,
then the matrix $\sfoldmatrix_r$ has the form:
\begin{equation}\label{eq:matrix_diag}
\begin{bmatrix}
           &          &        &         & \alpha+* &\\
           & \text{\Large*}&   &\alpha+* &          &\\
           &          & \reflectbox{$\ddots$} & &   &\\
           & \alpha+* &        & \text{\Large*}&    &\\
\alpha + * &          &        &         &          &\\
\end{bmatrix}.
\end{equation}
Since $ r\in \{2k-1\mid k=1,\ldots, B \} $ and $ i,j\in 
\{0,1\ldots,B\} $, we have
\begin{equation*}
\begin{split}
& |r+2(i+j)| = r+2B \Rightarrow   i+j = B,\\
& |r+2(i-j)| = r+2B  \Rightarrow  i=B, j=0 \text{ or } 
i=0, j=B.
\end{split}
\end{equation*}
Therefore, either  $ |r+2(i+j)| = r+2B $ or $ |r+2(i-j)| = 
r+2B $ implies $ i+j=B $, so $ \eval_{r+2B} $ only 
appears on the anti-diagonal of the matrix $ \sfoldmatrix_r 
$. Conversely, every element on the anti-diagonal of $ 
\sfoldmatrix_r $ is equal to $\eval_{r+2B}+\eval_{|r+2(i-j)|} $ 
for some $ i,j\in \{0, 1,\ldots,B\} $. Thus $\sfoldmatrix_r$ has the
form \eqref{eq:matrix_diag} and its determinant is a univariate 
polynomial of degree $ B+1 $ in $ \alpha $.

Now we consider the case that $ \eval_{r} $ is substituted 
by $ \alpha $. 
Similarly, because $ r\in \{2k-1\mid k=1,\ldots, B \} $  and $ i,j\in 
\{0,1\ldots,B\} $, we have 
\begin{equation}\label{eqn:locate_a_r}
\begin{split}
& |r+2(i+j)| = r \Rightarrow i=j=0, \\
& |r+2(i-j)| = r \Rightarrow i=j \text{ or } i=j-r \text{ if } j\geq r.
\end{split}
\end{equation}
Therefore, if $ r>B$ then $i=j$ in \eqref{eqn:locate_a_r}, so 
$\eval_r $ only appears on the main diagonal of 
$ \sfoldmatrix_r $. On the other hand,  every element on the 
main diagonal of $ \sfoldmatrix_r $ is equal to 
$ \eval_{|r+2(i+i)|} + \eval_r $ for some $ 
i\in \{0,1, \ldots, t\} $. Hence, if $ r>B$ then the 
determinant of $ \sfoldmatrix_r $ is a polynomial of degree 
$ B+1 $ in $ \alpha $. Assume that $ r\leq B $. From 
\eqref{eqn:locate_a_r}, we see that after 
substituting $ \eval_r $ by $ \alpha $, the matrix $ 
\sfoldmatrix_r $ has the form:
\begin{equation}\label{eq:matrix_antidiag}
\begin{bmatrix}
\alpha + * & \cdots & \alpha + * &       &    *       &  \\
           & \ddots &            &\ddots &            &  \\
           &        & \ddots     &       & \alpha + * &\\
           &   *    &            &\ddots & \vdots     &\\
           &        &            &       &\alpha+*    &\\

\end{bmatrix}.
\end{equation}
According to Lemma \ref{lem:2det_neq_0}, the determinant of 
the matrix \eqref{eq:matrix_antidiag} is a univariate 
polynomial of degree $ B+1 $ in $ \alpha $. 
\end{IEEEproof} 

\begin{example}
For $B=3$, we have $3B=9$ evaluations
$\{\eval_{2i-1}\}_{i=1}^{3B}$ obtained from the black box
for $f$ at inputs
$\gamma_{i}=(\omega^{2i-1}+\omega^{-(2i-1)})/2$.
We construct the following $6\times 4$ matrix:
        \begin{equation*}
        \sfoldmatrix = 
	\begin{bmatrix}
        2 \eval_1 & \eval_3 + \eval_1 & \eval_5 + \eval_3 & 
        \eval_7 + \eval_5 \\
        2 \eval_3 & \eval_5 + \eval_1 & \eval_7 + \eval_1 & 
        \eval_9 + \eval_3 \\
        2 \eval_5 & \eval_7 + \eval_3 & \eval_9 + \eval_1 & 
        \eval_{11} + \eval_1 \\
        2 \eval_7 & \eval_9 + \eval_5 & \eval_{11} + \eval_3 & 
        \eval_{13} + \eval_1 \\
	2 \eval_9 & \eval_{11} + \eval_7 & \eval_{13} + \eval_5 & 
        \eval_{15} + \eval_3 \\
	2 \eval_{11} & \eval_{13} + \eval_9 & \eval_{15} + \eval_7 & 
        \eval_{17} + \eval_5 \\
        \end{bmatrix}\in \KK^{6\times 4}.
        \end{equation*}
For $r=1,3,5$, the matrices $\sfoldmatrix_r$ are $4\times 4$
submatrices of the matrix $G$.
The matrix $\sfoldmatrix_1$ consists of the first $4$ rows
of $\sfoldmatrix$. If we substitute $\eval_1$ or $\eval_7$
by a symbol $\alpha$, then the determinant of 
$ \sfoldmatrix_1 $ is univariate polynomial of degree $4$
in $\alpha$. The matrix $\sfoldmatrix_3$ consists of the
second to the fifth row of $\sfoldmatrix$ and the
determinant of $\sfoldmatrix_3$ becomes a univariate
polynomial of degree $4$ in $\alpha$ if $\eval_3$ or
$\eval_9$ is substituted by $\alpha$. Similarly, the matrix
$\sfoldmatrix_5$ consists of the last $4$ rows of $G$.
Substituting $\eval_5$ or $\eval_{11}$ by $\alpha$,
$\det (\sfoldmatrix_5)$ is a univariate polynomial of degree
$4$ in $\alpha$.

Suppose there is one error $\eval_{2\ell-1}\neq f(\gamma_{2\ell-1})$ 
in the $3B$ evaluations. Here is how we correct this single
error for all possible $\ell$'s:
{\setlength{\leftmargini}{\wd\conditionone} 
\begin{enumerate}[(1)]
\item if $\ell\in \{1,2,3\}$, then substitute
$\eval_{2\ell-1}$ by $\alpha$ and compute the roots of
$\det(\sfoldmatrix_{2\ell-1})$, and the roots are
candidates for $f(\gamma_{2\ell-1})$;
\item if $\ell\in \{4,5,6\}$, then substitute
$\eval_{2\ell-1}$ by $\alpha$ and compute the roots of
$\det(\sfoldmatrix_{2(\ell-3)-1})$, and the roots are 
candidates for $f(\gamma_{2\ell-1})$;
\item if $\ell\in \{7,8,9\}$, then $f(x)$ can be recovered
by applying Prony's algorithm on the sequence 
$(\eval_{2i-1})_{i=-5}^{6}$. 
\end{enumerate}
}

\end{example}

We summarize the process of correcting one
error from $3B$ evaluations
in Algorithm~\ref{alg:cheb_1error} below. 
The Algorithm~\ref{alg:cheb_1error}
returns a list of sparse 
interpolants containing $f(x)$  
if there is at most one error in the $3B$ 
evaluations of $f(x)$. 
Also, $f(x)$ is not  
distinguishable from other interpolants (if there are any) 
in the list, because all interpolants returned by the 
algorithm satisfy the output conditions and $f(x)$ could be  
any one of them.   
However, if in the Algorithm~\ref{alg:cheb_1error} we have 
$\KK=\RR$, $\omega_{\sigma}>1$, $3B\geq 2B+2E=2B+2$ and the 
$3B$ evaluations contain at most one error,  
then $f(x)$ will be the only  
interpolant in the output. 
This is explicitly stated in 
\citeP[Corollary~2.4]{ArKa15}, 
which is a consequence of   
a generalization of Descartes's  
rule of signs to orthogonal polynomials by 
Obrechkoff's theorem \citeP[Theorem 1.1]{DiRa2009}. 
The Algorithm~\ref{alg:cheb_1error} will return  
FAIL if 
no sparse interpolants satisfy the output 
conditions.  

\vspace{-1ex} 
\begin{algorithm}
{\itshape A list-interpolation algorithm for Chebyshev-1
sparse polynomials with evaluations containing at most one
error.}
\newline
\ruleabovetwolineheader 
\label{alg:cheb_1error}
{\itshape Input:\/}
\begin{itemize}
\item[\ourtriang]
A black box representation of a polynomial $ f\in \KK[x]
$ where $ \KK $ is a field of 
scalars with 
characteristic $ \neq 2 $ and $f$ is  
a linear combination of \chebT
polynomials. The black box for $f$
returns the same (erroneous) output when 
probed 
multiple times at the same input.

\item[\ourtriang]
An upper bound $B$ of the sparsity of $f$.

\item[\ourtriang]
An upper bound $D$ of the degree of $f$.

\item[\ourtriang]
$ \omega\in \KK\setminus\{0\} $ has order $\geq 4D+1$. 

\item[\ourtriang]
An algorithm that computes all roots $\in \KK$ of a polynomial $\in \KK[x]$.
\end{itemize}

{\itshape Output:\/}
\begin{itemize}
\item[\ourtriang]
Either a list of sparse polynomials $\{ \fsupbr{1}, \ldots,
\fsupbr{M} \}$ with each $\fsupbr{k}$ 
$(1\leq k\leq M)$ satisfying:

\begin{itemize}
\item[\ourtriang]
$\fsupbr{k}$ has sparsity $ \leq B $ and degree $\leq D$;

\item[\ourtriang]
$\fsupbr{k}$ is represented by its \chebT 
term degrees and coefficients;

\item[\ourtriang]
there is at most one 
index $i\in \{1,2,\ldots, 3B\} $ such
that $\fsupbr{k}(\gamma_{2i-1})\neq \eval_{2i-1}$ where 
$\gamma_i = (\omega^{2i-1}+\omega^{-(2i-1)})/2$ and 
$\eval_{2i-1}$ is the output of the black box 
probed at input $\gamma_{2i-1}$;

\item[\ourtriang]
$f$ is contained in the list,
\end{itemize}
\item[\ourtriang]
or FAIL. 
\end{itemize}


{\setlength{\leftmargini}{\myleftmargini}

\begin{enumerate}[Step 1:]
\setlength{\itemindent}{\myitemindentstep}

\item
{\itshape For $ i=1,2,\ldots,3B $, get the
	 output $ \eval_i $ of the black box for $f$ at input 
	 $\gamma_i = (\omega^{2i-1}+\omega^{-(2i-1)})/2$.
	Let $ L $ be an empty list.}

\item \label{step:error_at_tail_cheby} 
{\itshape 
Use Algorithm~\ref{alg:prony} on the 
sequence $(\eval_{2i-1} )_{i=-(2B-1)}^{2B}$. 
If Algorithm~\ref{alg:prony} returns 
a polynomial of the following form: 
$\sum_{j=1}^t \frac{c_j}{2} 
(\omega^{-\delta_j}x^{2\delta_j}+\omega^{\delta_j}x^{-2\delta_j})$ with $c_j\in \KK, \; t\leq B,
\delta_j\leq D$, then let $\bar{f} = 
\sum_{j=1}^t c_j T_{\delta_j}(x)$.
If 
there is at most one 
index $i\in\{1,\ldots, 3B\}$ 
such that $\bar{f}(\gamma_{2i-1})\neq \eval_{2i-1}$, 
then add $\bar{f}$ to the list $L$.
}

Step~\ref{step:error_at_tail_cheby} will add $f$ to the list $L$ if 
the error is in $\{\eval_{2i-1}\}_{i=2B+1}^{3B}$.
 



\item \label{step:error_at_head_cheby}
{\itshape For $ \ell = 1, \ldots, B $,  }

\begin{enumerate}[{\theenumi}(a):]
\setlength{\itemindent}{\myitemindentonea}

	\item {\itshape substitute $ \eval_{2\ell-1} $ 
	by a symbol $\alpha $ in the matrix 
	$ \sfoldmatrix_{2\ell-1} $; compute the determinant 
	of $ \sfoldmatrix_{2\ell-1} $ and denote it by
	$ \Delta_{2\ell-1}(\alpha) $; }

According to Lemma \ref{lem:foldmatrix_detneq0}, $ 
\Delta_{2\ell-1}(\alpha) $ is a univariate polynomial of degree 
$ B+1 $ in $ \alpha $.
	
	\item \label{step:comp_zeta_cheby} 
	{\itshape
	compute all solutions of the equation 
	$ \Delta_{2\ell-1}(\alpha ) = 0 $ in $ \KK $; denote the 
	solution set as $\{ \xi_1,\ldots,\xi_b \} $;}

	\item {\itshape for $ k=1,\ldots,b $, }

	\begin{enumerate}[{\theenumi}(\theenumii)i:]
\setlength{\itemindent}{\myitemindentoneaiii}

        \item {\itshape substitute $ \eval_{2\ell-1} $ by $
                \xi_k $;}
        \item {\itshape use Berlekamp/Massey algorithm to compute the
                the minimal linear generator of the new sequence
                $(\eval_{2i-1})_{i=-3B+1}^{3B} $ and
                denote it by $\Lambda(z)$;}
        \item {\itshape if $\deg (\Lambda(z)) \leq 2B$, repeat Step
                \ref{step:error_at_tail_cheby}.}
        \end{enumerate}

%
%

\end{enumerate}

If the error is $\eval_{2\ell-1}$ with $1\leq \ell\leq B$, 
that is $\eval_{2\ell-1}\neq f(\gamma_{2\ell-1})$, then 
we substitute $\eval_{2\ell-1}$ by a symbol $\alpha$. As 
the correct value $f(\gamma_{2\ell-1})$ is a solution of 
$\Delta_{2\ell-1}(\alpha)=0$, that is 
$f(\gamma_{2\ell-1})=\xi_k$ 
for some $ k\in \{1,\ldots,b\} $, Step 
\ref{step:error_at_head_cheby} will add $f$ into the list 
$L$.

\item \label{step:error_at_middle_cheby}
{\itshape For $ \ell = B+1, \ldots, 2B $,  }

\begin{enumerate}[{\theenumi}(a):]
\setlength{\itemindent}{\myitemindentonea}

	\item {\itshape substitute $ \eval_{2\ell-1} $ 
	by a symbol $\alpha $ in the matrix 
	$ \sfoldmatrix_{2(\ell-B)-1} $;
	compute the determinant of $ \sfoldmatrix_{2(\ell-B)-1} $ 
	and denote it by $ \Delta_{2\ell-1}(\alpha) $; }
	
	According to Lemma \ref{lem:foldmatrix_detneq0}, $ 
	\Delta_{2\ell-1}(\alpha) $ is a univariate polynomial 
	of degree $ B+1 $ in $ \alpha $.
	
	\item \label{step:comp_zeta_cheby_midd} 
	{\itshape
		compute all solutions of the equation 
		$ \Delta_{2\ell-1}(\alpha ) = 0 $ in $ \KK $; 
		denote the solution set as 
		$\{ \xi_1,\ldots,\xi_{b'} \} $;}
       
	\item {\itshape for $ k=1,\ldots,b' $, }

        \begin{enumerate}[{\theenumi}(\theenumii)i:]
\setlength{\itemindent}{\myitemindentoneaiii}

        \item {\itshape substitute $ \eval_{2\ell-1} $ by $
                \xi_k $;}
        \item {\itshape use Berlekamp/Massey algorithm to compute the
                the minimal linear generator of the new sequence
                $(\eval_{2i-1})_{i=-3B+1}^{3B} $ and
                denote it by $\Lambda(z)$;}
        \item {\itshape if $\deg( \Lambda(z)) \leq 2B$, repeat Step
                \ref{step:error_at_tail_cheby}.}
        \end{enumerate}
	
	%
	%
	
\end{enumerate}

If the error is $\eval_{2\ell-1}$ $(B+1\leq \ell \leq 
2B)$, that is $\eval_{2\ell-1}\neq f(\gamma_{2\ell-1})$, 
we also substitute $\eval_{2\ell-1}$ by a symbol $\alpha$. 
As the solution set $\{ \xi_1,\ldots,\xi_{b'} \} $ of 
$ \Delta_{2\ell-1}(\alpha ) = 0 $ contains $f(\gamma_{2\ell-1})$,
Step \ref{step:error_at_middle_cheby} will add $f$ into the list 
$L$.

\item 
{\itshape If the list $L$ is empty, then return FAIL, otherwise return
the list $L$. }
\rulebelowtext{0.8ex}
\end{enumerate}
}
\end{algorithm}


\begin{prop}\label{prop:cheb_outputbound_1error}
	The output list of Algorithm \ref{alg:cheb_1error} 
	contains $\leq 2B^2+2B+1 $ polynomials.
\end{prop}
\begin{IEEEproof} 
	The Step \ref{step:error_at_tail_cheby} in Algorithm 
	\ref{alg:cheb_1error} produces $ \leq 1 $ polynomial, 
	and both Step \ref{step:error_at_head_cheby} and Step 
	\ref{step:error_at_middle_cheby} produce $ \leq B(B+1) 
	$ polynomials. Hence the final output list has $\leq 
	1+2B(B+1) $ polynomials.
\end{IEEEproof} 

\subsection{Correcting $E$ Errors}
\label{subsec:cheb_Eerrors}
The settings for $ f(x) $ are the same as in Section 
\ref{subsec:cheb_1error}.
We show how to list interpolate $f(x)$ from $\numofevals$ 
evaluations containing $\leq E $ errors, where
\begin{equation}\label{eqn:cheb_numofevals}
\numofevals = \left\lfloor \frac{3}{2}E + 2\right\rfloor B.
\end{equation}
Let $\numofblocks = \left\lfloor E/2 \right\rfloor$. Choose 
$\omega_{1},\ldots,\omega_{\numofblocks},\omega_{\numofblocks
+ 1} \in \KK\setminus\{0\}$ such that $\omega_\sigma$ has
order $\geq 4D+1$ for $1\leq \sigma \leq \numofblocks+1$.

If $E$ is even then $\numofevals = (E/2)3B + 2B$. 
The problem is reduced to 
one of the 
following situations: 
(1) the last block $ (\eval_{\numofblocks+1,2i-1} )_{i=1}^{2B} $ 
of length $2B$ is free of errors, 
or (2) there is some block $(\eval_{\sigma,2i-1})_{i=1}^{3B}$
with $1\leq \sigma\leq  E/2 $
of length $3B$ that contains at most $1$ error. 
These two situations can be 
handled 
by the 
Algorithm~\ref{alg:prony} and Algorithm~\ref{alg:cheb_1error}, respectively. 

If $E$ is odd then $E= 2\cdot \numofblocks + 1$ and 
$\numofevals =( \numofblocks+1 ) 3B $. 
Thus, there is some block 
$( \eval_{\sigma,1},\ldots, \eval_{\sigma,3B} )$ 
with $1\leq \sigma\leq \numofblocks+1$  
of length $3B$ that contains at most $1$ error;  
we can use
the Algorithm~\ref{alg:cheb_1error} on this block to list 
interpolate $f(x)$.

\begin{remark}\label{rmk:cheb_outputbound_Eerrors}
	For every $\sigma\in \{1,\ldots, \left\lfloor 
	E/2 \right\rfloor \}  $, we apply Algorithm 
	\ref{alg:cheb_1error} on the block 
    $(\eval_{\sigma,2i-1})_{i=1}^{3B}$    
    which will result in $\leq \left\lfloor E/2 
	\right\rfloor (2B^2+2B+1) $ 
	polynomials by Proposition 
	\ref{prop:cheb_outputbound_1error}. 
	The length of the last block depends on the value of $ 
	E $, and we 
	have the following 
	different upper bounds on 
	the number of resulting polynomials:
{\setlength{\leftmargini}{\wd\conditionone} 
	\begin{enumerate}[(1)]
		\item $ (E/2) (2B^2+2B+1 )+1$, if $E$
		is even;
		\item $\left(\left\lfloor E/2 \right\rfloor 
		+ 1\right)(2B^2+2B+1)$, if $ E $ is odd.
	\end{enumerate}
}
\end{remark}

Due to Obrechkoff's theorem
\citeP[Theorem 1.1]{DiRa2009}, 
a generalization of Descartes's 
rule of signs to orthogonal polynomials, 
our approach for
correcting $E$ errors gives a unique valid sparse interpolant when
$\KK=\RR$, $N\geq 2B+2E$ and $\omega_{\sigma}>1$ 
\citeP[Corollary~2.4]{ArKa15}. 
Similar to the case of standard power 
basis, 
if $N<2B+2E$ then
there can be $\geq 2$ valid sparse interpolants in \chebT basis as
shown by the following example.

\begin{example}\label{example:cheb_list}
Choose $\omega > 1$. The polynomials $h$, $\fsupbr{1}$
and $\fsupbr{2}$, given in Example \ref{example:power_list}, can be
represented in \chebT basis  
using the following formula \citeP[P. 303]{Fraser1965}, 
\citeP[P. 412]{Cody1970}, 
\citeP[Eq. (2)]{Mathar2006}:
\begin{equation}\label{eq:power_to_cheb}
x^d =
\frac{1}{2^{d-1}}
\sum_{\substack{j=0\\d-j \text{ is even }}}^{d}
\binom{d}{(d-j)/2}
\times \left\{
\begin{tabular}{ll}
$T_{j}(x)$ & if $j \ge 1$,
\\
$\frac{1}{2}$ & if $j = 0$.
\end{tabular}
\right.
\end{equation}
Moreover, the formula
\eqref{eq:power_to_cheb} implies that $\fsupbr{1}$ is a
linear combination of the odd degree \chebT polynomials 
$T_{2j-1}(x)$ $(j=1,2,\ldots, B)$, 
and $\fsupbr{2}$ is a linear combination of the
even degree \chebT polynomials $T_{2j-2}(x)$ $(j=1,2,\ldots, B)$,
which means both $\fsupbr{1}$ and $\fsupbr{2}$ have sparsity $\leq B$
in \chebT basis as well. Therefore, $\fsupbr{1}$ and $\fsupbr{2}$
are also valid interpolants in \chebT basis for the
$2B+2E-1$ evaluations given in \eqref{eq:overreal_evals} (if we 
assume $B$ is an upper bound on the sparsity of the black-box 
polynomial $f$ and $E$ is an upper bound on the number of errors 
in the evaluations). 

Again, we remark that one of the valid
interpolants, $\fsupbr{1}$ and $\fsupbr{2}$, must have sparsity $B$ 
since otherwise uniqueness is a consequence of the Obrechkoff's theorem 
\citeP[Theorem 1.1]{DiRa2009}. In this example, $h$ also has $2B$
terms in \chebT basis because $\deg(h)=2B-1$ and $h$ has $2B-1$ real 
roots $\omega^i>1,\; i=1,\ldots, 2B-1$. Thus both 
$\fsupbr{1}$ and $\fsupbr{2}$ have sparsity $B$ in \chebT basis.
One can also make 
$h$, $\fsupbr{1}$ and $\fsupbr{2}$ sparse 
with respect to the \chebT 
basis by the following substitutions:
\[
x = T_k(y), \; \omega = T_k(\bar{\omega}) \text{ for some } k\gg 1. 
\]
\end{example}

For $\KK=\CC$, we usually choose $\omega$ as a root of unity.
But then we may need $2B(2E+1)$ evaluations to get a unique
interpolant. Here is an example from \citeP[Theorem 3]{KaPe14}, simply
by changing the power basis to Chebyshev-1 basis.

\begin{example}\label{example:cheb_unique_decoding}
Consider the following two polynomials:
\begin{align*}
& f_1(x) = \frac{1}{t}\sum_{j=0}^{t-1} T_{2j\frac{m}{2t}}(x), \\ 
& f_2(x) = -\frac{1}{t}\sum_{j=0}^{t-1} T_{(2j+1)\frac{m}{2t}}(x),
\end{align*}
where $m\geq 2t(2E+1)-1$ and $2t$ divides $m$. Let $\omega$ be a
primitive $m$-th root of unity. 
Let 
\[
b = (\underbrace{0,\ldots,0}_{t-1}, 1, \underbrace{0, \ldots, 0}_{t-1})
\in \KK^{2t-1}.
\]
The evaluations of $f_1$ at $\frac{\omega^i + \omega^{-i}}{2}$ for
$i=1,2,\ldots, 2t(2E+1)-1$ are
\[
(\underbrace{b,\; 1, \ldots, b,\; 1}_{2E \text{ pairs of }
(b,\; 1)},  
b)
\in \KK^{2t(2E+1)-1}.
\]
The evaluations of $f_2$ at $\frac{\omega^i + \omega^{-i}}{2}$ for
$i=1,2,\ldots, 2t(2E+1)-1$ are
\[
(\underbrace{b, -1, \ldots, b, -1}_{2E \text{ pairs of } 
(b,-1)},  
b)
\in \KK^{2t(2E+1)-1}.
\]
Suppose we probe the black box for $f$ at 
$\frac{\omega^i + \omega^{-i}}{2}$ with
$i=1,2,\ldots, 2t(2E+1)-1$ sequentially, and obtain the following sequence of
evaluations:
\begin{equation*}
\eval=(\underbrace{b,\; 1,\ldots, b,\; 1}_{E \text{ pairs of
} (b,\; 1) }, 
\underbrace{b, -1, \ldots, b, -1}_{E \text{ pairs of } 
(b, -1)},  
b)\in
\KK^{2t(2E+1)-1}
\end{equation*}
Assume $B=t$ and there are $E$ errors in the sequence $\eval$.
Then both $f_1$ and $f_2$ are valid interpolants for $\eval$.
More specifically, $f_1$ is a valid interpolant for $\eval$ if the $E
$ errors are $\eval_{2t}, \eval_{2t\cdot 2}, \ldots, 
\eval_{2t\cdot E}$; $f_2$ is a valid interpolant for $\eval$ if the $E
$ errors are $\eval_{2t(E+1)}, \eval_{2t(E+2)}, \ldots, \eval_{2t\cdot
2E}$.
\end{example}

\begin{remark}\label{remark:chebyshev_234} 
Polynomials in Chebyshev-2, Chebyshev-3 and Chebyshev-4 bases can be
transformed into Laurent polynomials using the formulas
given in \citeP[Sec. 1, (7)-(9)]{IKY18}. Therefore, our
approach to list-interpolate black-box polynomials in
Chebyshev-1 bases also works for black-box polynomials in 
Chebyshev-2, Chebyshev-3 and Chebyshev-4 bases.
\end{remark}

\bibliographystyle{IEEEtran}

\bibliography{%
error_correction%
}
\begin{IEEEbiographynophoto}{Erich L. Kaltofen}
received his Ph.D. degree in Computer Science in 1982
from Rensselaer Polytechnic Institute. He was an Assistant Professor
of Computer Science at the University of Toronto and an Assistant,
Associate, and Professor at Rensselaer Polytechnic Institute. Since
1996 he is a Professor of Mathematics at North Carolina State
University, since 2018 he is also an Adjunct Professor
of Computer Science at Duke University.
He has held visiting positions at ENS Lyon, MIT,
Pierre and Marie Curie U., 
Paris, U. Grenoble, and U. Waterloo.  His research interests
are theoretical aspects of symbolic computation and its application
to computational problems.  Kaltofen was the Chair of ACM's
Special Interest Group on Symbolic \& Algebraic Manipulation 1993-95.
In 2009 Kaltofen was selected an ACM Fellow.
\end{IEEEbiographynophoto}

\begin{IEEEbiographynophoto}{Zhi-Hong Yang}
received her Ph.D. degree in Applied Mathematics in 2018
from Chinese Academy of Sciences,  under the advisement of
Lihong Zhi. 
Her Ph.D. thesis is on the complexity of computing real radicals of polynomial ideals. 
During 2018--2020, she was 
a postdoctoral scholar in 
the Mathematics Department at North Carolina State University, and a
postdoctoral visitor in the Computer Science Department at Duke
University 
Currently,
she is an associate researcher
in the College of Mathematics and Statistics
at Shenzhen University. She is working on
on polynomial interpolation with error correction and its applications.
\end{IEEEbiographynophoto}

\vspace{6ex} 

\appendix
\section{Appendix}

\newsavebox{\xixi}\savebox{\xixi}{$\xi_{1,i}, \xi_{2,i}\;$}
\newlength{\secondcolwidth}\setlength{\secondcolwidth}{\columnwidth}
\addtolength{\secondcolwidth}{-\wd\xixi}
\addtolength{\secondcolwidth}{-0.4em}

\tablefirsthead{\hline}
\tablehead{\hline\multicolumn{2}{|@{\hspace*{0.2em}}l@{\hspace*{0.2em}}|}
{{\bfseries Notation continued} (in alphabetic order):}\\\hline}
\tabletail{\hline\multicolumn{2}{|r|}{\itshape Continued on next page} \\\hline}
\tablelasttail{}
\setlongtables
\begin{supertabular}{%
|%
@{\hspace*{0.2em}}%
l%
@{}p{\secondcolwidth}%
@{\hspace*{0.2em}}
|%
}
\multicolumn{2}{|@{\hspace*{0.2em}}l@{\hspace*{0.2em}}|}
{{\bfseries Notation} (in alphabetic order): }
\\ \hline
$\eval_i$&the output of the black box for $f$ at input $\omega^i$ 
\\
$\alpha$&a symbol that substitute the single error in a block of $3B$
outputs of the black box for $f$
\\
$\alpha_1, \alpha_2$ 
&symbols that substitute the two errors in a
block of $4B$ outputs of the black box for $f$
\\
$B$&$\geq t$, an upper bound on the sparsity of $f$
\\
$b$&number of solutions to polynomial equation(s) for hypothetical errors
\\
$\beta$&$=(\omega+1/\omega)/2$, evaluation point of \chebT polynomials
\\
$c_j$&the coefficient 
of the $j$-th term of $f$
\\
$D$&$\geq |\delta_j|$, an upper bound on the absolute values of the degree of $f$
\\
$\delta_j$&the $j$-th term degree of $f$
\\
$\Delta$&a matrix determinant 
\\
$E$&an upper bound on the number of errors that is input to the algorithm
\\
$f$&the black-box polynomial
\\
$\gamma_{i}$&$=(\omega^i+1/\omega^i)/2$, inputs of
the black box for $f$ if $f$ is in Chebyshev bases
\\
$\sfoldmatrix_r$&$\in \KK^{(B+1)\times(B+1)}$, the Hankel$+$Toeplitz
matrix with $ \eval_{|r+2(i+j)|} + \eval_{|r+2(i-j)|}$ on its $(i+1)$-th
row and $(j+1)$-th column
\\
$H_r$&$\in \KK^{(B+1)\times(B+1)}$, the Hankel matrix
with $\eval_{r+i-1}, \eval_{r+i}, \ldots, \eval_{r+i-1+B}$
on its $i$-th row 
\\
$\KK$&a field of 
scalars with 
characteristic $\neq 2$  
\\
$\xi_i$&candidates for the correct value $f(\omega^\ell)$ if
$\eval_\ell$ is assumed to be an error
\\
$\xi_{1,i}, \xi_{2,i}\;$
&candidates for the pair of correct values
$f(\omega^{\ell_1})$, $f(\omega^{\ell_2})$ if $\eval_{\ell_1}$ and
$\eval_{\ell_2}$ are assumed to be errors
\\
$\ell$&the error location in the outputs of the black box
for $f$ if $E=1$
\\
$\ell_1,\ell_2$&the error locations in the outputs of the
black box for $f$ if $E=2$
\\
$L$&the output list of our list decoding algorithms
\\
$\Lambda$&the term locator polynomial 
\\
$M$&the number of the output polynomials of our
error-correcting algorithms
\\
$N$&the number of the evaluations by the black box for $f$
\\
$\omega$&a non-zero number in $\KK$, evaluation base point for the
black-box polynomial $f$ when only one block of evaluations are needed
\\
\hbox to \wd\xixi{$\omega_\sigma$\hss}
&$\sigma=1,2,\ldots,\numofblocks+1$, non-zero numbers
in $\KK$, evaluation base points for the black box polynomial $f$
when multiple blocks of evaluations are needed
\\
$\rho_j$&$1\leq j\leq t$, the roots of the term locator
polynomial $\Lambda$
\\
$t$&the actual number of terms of $f$
\\
$\numofblocks$&$=\lfloor E/3 \rfloor$ if the black-box polynomial $f$ 
is in power basis, or $=\lfloor E/2 \rfloor$ if the black-box
polynomial $f$ is in Chebyshev bases
\\ 
$\zeta_i$&distinct, algorithm-dependent arguments in $\KK$ 
\\
\hline
\end{supertabular}


\end{document}